\documentclass[twocolumn,hyperpdf,amsmath,amssymb,aps,prd,10pt,superscriptaddress,nofootinbib,noeprint,preprintnumbers,floatfix]{revtex4-1}

\usepackage{graphicx, color}

\usepackage[letterspace=-10]{microtype} % gives access to \lsstyle which will tighten letter spacing

\usepackage{bm, amsmath, amsfonts}
\usepackage{multirow, tabularx, dcolumn}

\usepackage[utf8]{inputenc} %for special characters in bibtex
\usepackage{hyperref}
\definecolor{jlab_red}{RGB}{192,39,45}
\definecolor{jlab_orange}{RGB}{249,102,0}
\definecolor{jlab_blue}{RGB}{47,122,121}
\definecolor{jlab_green}{RGB}{65,125,10}

\newcommand{\KKbar}[0]{$K\overline{K}$ }

\hypersetup{%
pdftitle = {Isoscalar $\pi\pi, K\overline{K}, \eta\eta$ scattering and the $\sigma, f_0, f_2$ mesons from QCD},
pdfsubject = {QCD},
pdfkeywords = {QCD, Hadron, Physics, JLab, Lattice, Meson, Scattering},
pdfauthor = {Hadron Spectrum Collaboration},
colorlinks = {true},
filecolor = {black},
linkcolor = {jlab_blue},
menucolor = {black},
citecolor = {jlab_green},
urlcolor = {jlab_green},
}{}

\begin{document}

\preprint{JLAB-THY-17-2534}

\title{Isoscalar $\pi\pi, K\overline{K}, \eta\eta$ scattering and the $\sigma, f_0, f_2$ mesons from QCD}

%%%%%%%%%%%%%%%%%%%%%%%%%%%%%%%%%%%%%%%%%%%%%%%%%%%%%%%%%%%%%%%%%%%%%%
\author{Raul~A.~Brice\~{n}o}
\email{briceno@jlab.org}
\affiliation{\lsstyle Thomas Jefferson National Accelerator Facility, 12000 Jefferson Avenue, Newport News, VA 23606, USA}
\affiliation{\lsstyle Department of Physics, Old Dominion University, Norfolk, VA 23529, USA}
%%%%%%%%%%%%%%%%%%%%%%%%%%%%%%%%%%%%%%%%%%%%%%%%%%%%%%%%%%%%%%%%%%%%%%
\author{Jozef~J.~Dudek}
\email{dudek@jlab.org}
\affiliation{\lsstyle Thomas Jefferson National Accelerator Facility, 12000 Jefferson Avenue, Newport News, VA 23606, USA}
\affiliation{Department of Physics, College of William and Mary, Williamsburg, VA 23187, USA}
%%%%%%%%%%%%%%%%%%%%%%%%%%%%%%%%%%%%%%%%%%%%%%%%%%%%%%%%%%%%%%%%%%%%%%
\author{Robert~G.~Edwards}
\email{edwards@jlab.org}
\affiliation{\lsstyle Thomas Jefferson National Accelerator Facility, 12000 Jefferson Avenue, Newport News, VA 23606, USA}
%%%%%%%%%%%%%%%%%%%%%%%%%%%%%%%%%%%%%%%%%%%%%%%%%%%%%%%%%%%%%%%%%%%%%%
\author{David~J.~Wilson}
\email{djwilson@maths.tcd.ie}
\affiliation{School of Mathematics, Trinity College, Dublin~2, Ireland}
%%%%%%%%%%%%%%%%%%%%%%%%%%%%%%%%%%%%%%%%%%%%%%%%%%%%%%%%%%%%%%%%%%%%%%

\collaboration{for the Hadron Spectrum Collaboration}
\date{\today}
%\pacs{}

%%%%%%%%%%%%%%%%%%%%%%%%%%%%%%%%%%%%%%%%%%%%%%%%%%%%%%%%%%%%%%%%%%%%%%
\begin{abstract}
We present the first lattice QCD study of coupled isoscalar $\pi\pi,K\overline{K},\eta\eta$ $S$- and $D$-wave scattering extracted from discrete finite-volume spectra computed on lattices which have a value of the quark mass corresponding to $m_\pi\sim391$~MeV. In the $J^P=0^+$ sector we find analogues of the experimental $\sigma$ and $f_0(980)$ states, where the $\sigma$ appears as a stable bound-state below $\pi\pi$ threshold, and, similar to what is seen in experiment, the $f_0(980)$ manifests itself as a dip in the $\pi\pi$ cross section in the vicinity of the $K\overline{K}$ threshold. For $J^P=2^+$ we find two states resembling the $f_2(1270)$ and $f_2'(1525)$, observed as narrow peaks, with the lighter state dominantly decaying to $\pi\pi$ and the heavier state to $K\overline{K}$. The presence of all these states is determined rigorously by finding the pole singularity content of scattering amplitudes, and their couplings to decay channels are established using the residues of the poles. 
\end{abstract}
%%%%%%%%%%%%%%%%%%%%%%%%%%%%%%%%%%%%%%%%%%%%%%%%%%%%%%%%%%%%%%%%%%%%%%

\maketitle

%%%%%%%%%%%%%%%%%%%%%%%%%%%%%%%%%%%%%%%%%%%%%%%%%%%%%%%%%%%%%%%%%%%%%%
%%%%%%%%%%%%%%%%%%%%%%%%%%%%%%%%%%%%%%%%%%%%%%%%%%%%%%%%%%%%%%%%%%%%%%
%%%%%%%%%%%%%%%%%%%%%%%%%%%%%%%%%%%%%%%%%%%%%%%%%%%%%%%%%%%%%%%%%%%%%%

%%%%% INTRO %%%%%
\section{Introduction \label{Sec:introduction}}
% !TEX root = ../f0_paper.tex

%%%%% INTRO %%%%    

The composition of the lightest hadrons with scalar ($J^P =0^+$) quantum numbers remains a mystery. From one viewpoint, this is peculiar -- being rather light and thus having only limited possible decay channels, we might expect them to provide a relatively simple system to study, but yet they have proven to be rather challenging. The lightest \emph{isoscalar} scalar mesons are particularly interesting, as they appear to be two quite different objects -- an extremely broad and light $f_0(500)$ (historically called the $\sigma$, a name we will continue to use), and a rather narrow $f_0(980)$ appearing very close to the \KKbar threshold~\cite{Conforto:1967zz, Astier:1967zz}. These states appear in $\pi\pi$ scattering in a way which challenges our naive view of hadron resonances as peak-like enhancements in cross-sections, appearing rather as a very broad bump with a sharp \emph{dip} at the $K\overline{K}$ threshold~\cite{Protopopescu:1973sh, Hyams:1973zf, Grayer:1974cr, Estabrooks:1974vu, Kaminski:1996da, Chew:1960iv, Flatte:1976xv,Bargiotti:2003ev, Baru:2004xg, Ananthanarayan:2000ht, Colangelo:2001df, Zhou:2004ms, GarciaMartin:2011jx, Caprini:2005zr, Pelaez:2015qba}.

The $\sigma$ and $f_0(980)$ have commonly been placed together with the $K^\star_0(800)$ (or $\kappa$) and $a_0(980)$ resonances into a nonet of (broken) $SU(3)$ flavor. The $\kappa$ appears to be a strange analogue of the $\sigma$, being a very broad resonance enhancing $\pi K$ scattering at low energies, while the isovector $a_0(980)$ closely resembles the $f_0(980)$. The $a_0(980)$ appears at the $K\overline{K}$ threshold, and likely shares the $f_0(980)$'s strong coupling to $K\overline{K}$. It is observed through its decay to $\pi\eta$, but the lack of direct data on $\pi \eta$ elastic scattering limits the precision with which it can be studied.

In the simplest interpretation of the constituent quark model, the scalar nonet would arise from $q\bar{q}(^3\!P_0)$ constructions, but such an assignment for the states discussed above appears unnatural given that the related $q\bar{q}(^3\!P_{1,2})$ states lie at much higher masses. In addition, this picture provides no explanation for the near degeneracy of the the $f_0(980)$ and $a_0(980)$, nor why the $\sigma$ and $\kappa$ are apparently lighter than them. One suggestion which attempts to remedy this is to have the nonet be of dominantly $qq\bar{q}\bar{q}$ construction~\cite{Jaffe:1976ig}, with the $a_0(980)$ and $f_0(980)$ having hidden strangeness, making them heavier than the $\sigma$ which is proposed to contain only light quarks. Alternatively, given their proximity to the $K\overline{K}$ threshold, a natural suggestion is that the $a_0(980)$ and $f_0(980)$ might be dominated by $K\overline{K}$ molecular configurations, where the binding is provided by residual interhadron forces~\cite{Weinstein:1990gu} (see Ref.~\cite{Guo:2017jvc} for a review of resonances as hadronic molecules). Similarly, dynamical modeling which has a $q\bar{q}$ `seed' being dressed by strong coupling to its meson-meson decay modes has proven capable of generating light scalar resonances which resemble those in experiment~\cite{vanBeveren:1986ea,Tornqvist:1995kr,Boglione:2002vv,vanBeveren:2006ua}. Interpretation of the scalar mesons can be further complicated if glueball basis states are assumed to also play a role in the isoscalar case~\cite{Ochs:2006rb}.

In contrast to the scalar sector, the lightest isoscalar tensor mesons ($J^P=2^+$) present a situation that much more closely aligns with our na\"ive view of hadron resonances. In the constituent quark picture, two isoscalar states can be constructed from combinations of $u\overline{u}+d\overline{d}$ and $s\overline{s}$ pairs in the $^3\!P_2$ configuration. Experimentally the $f_2(1270)$ and $f_2'(1525)$ appear as ``bump-like" enhancements which are well described by Breit-Wigner parametrizations, with the lighter state dominantly decaying to the $\pi\pi$ final state ($84\%$), and the heavier state to $K\overline{K}$ ($89\%$)~\cite{Olive:2016xmw}. These decay characteristics are taken as support for the quark-model assignment $f_2(1270) \sim u\bar{u} + d\bar{d}$, $f_2'(1525) \sim s\bar{s}$ using the phenomenology of the `OZI' rule, where $q\bar{q}$ pair creation is proposed to dominate existing $q\bar{q}$ annihilation in hadron decays. The rates of two-photon decays of these states have also been presented as support for these assignments (see Ref.~\cite{Dai:2016ytz} and references therein).

The presence of a resonance of angular momentum, $J$, in a hadron scattering process has a rigorous signature in the form of a pole singularity in the partial-wave amplitude, $t_J(s)$. A pole at a complex value of Mandelstam $s = s_R = \big( m_R- i \tfrac{1}{2}\Gamma_R \big)^2$ influences scattering for real values of $s$ -- in particular for small value of the width, $\Gamma_R$, the effect for an isolated resonance is typically a narrow peak. In general, the scattering amplitude is a matrix in the space of kinematically open channels, and near a resonance pole the elements of the matrix take the form $t_{ij}\sim \frac{c_i\, c_j}{s_R-s }$, where $c_i$ is the coupling to the $i$ channel. A rigorous presentation of hadron resonances will take the form of a list of pole positions and the associated couplings.

As described above, dynamical models working at the level of constituent quarks and/or hadrons are informative, and offer frameworks within which experimental observations may be placed, but ultimately all observed hadron phenomena must have an origin within QCD. Lattice QCD provides an explicit numerical approach to studying QCD at energies relevant to resonance physics, where it is fundamentally non-perturbative. After discretizing the theory on a finite hypercubic grid, an ensemble of gluon field configurations can be generated, and using these, correlation functions evaluated. By computing appropriate two-point correlation functions, the discrete spectrum of QCD eigenstates in the finite-volume defined by the lattice can be determined. These can be related to scattering amplitudes through the L\"uscher formalism~\cite{Luscher:1986pf, Luscher:1990ux, Luscher:1991cf} and its extensions~\cite{Rummukainen:1995vs, He:2005ey, Christ:2005gi, Kim:2005gf, Guo:2012hv, Hansen:2012tf, Briceno:2012yi, Briceno:2014oea} which connect the volume dependence of the discrete spectrum to scattering amplitudes in an infinite volume.

An approach which has proven successful~\cite{Dudek:2014qha, Wilson:2014cna,Wilson:2015dqa,Dudek:2016cru} proceeds by parametrizing the energy-dependence of coupled-channel amplitudes and fitting a large set of energy levels, from one or more lattice volumes, within a kinematic window~\cite{Guo:2012hv}. A dense spectrum of energy levels will tightly constrain the possible energy-dependence of the scattering $t$-matrix, and to acquire as many energy levels as possible, we may consider systems with various total momenta. Currently this approach is limited to energies below three-hadron and higher thresholds -- to go beyond this an extension of this formalism is required, and progress in this direction is being made~\cite{Briceno:2017tce, Hammer:2017kms, Hansen:2014eka, Hansen:2015zga, Briceno:2012rv, Polejaeva:2012ut}.

In this first study of excited isoscalar meson resonances, we will calculate a version of QCD featuring light quarks which are heavier than those found experimentally, leading to a pion mass of 391 MeV. In this world, three- and four-meson thresholds appear at higher energies, providing a larger energy region in which we can perform rigorous extraction of scattering amplitudes. At this pion mass we have previously studied the other symmetry channels that would make up a flavor nonet: $I=1/2,\, S=\pm 1$ through coupled $\pi K, \eta K$ scattering~\cite{Dudek:2014qha,Wilson:2014cna}, and $I=1,\, S=0$ through coupled $\pi \eta, K\overline{K}$~\cite{Dudek:2016cru}. 

In Refs.~\cite{Dudek:2014qha,Wilson:2014cna}, the scattering matrix in the $\pi K, \eta K$ sector was found to be almost completely decoupled, with $\eta K \to \eta K$ scattering being rather weak, while the $\pi K$ $S$-wave was found to be attractive at threshold, and to feature no rapid energy dependence. This was interpreted as being due to a \emph{virtual bound-state} singularity over \mbox{200 MeV} below $\pi K$ threshold, as well as a much heavier broad scalar resonance pole above $1.3$ GeV lying far into the complex plane.

In the isovector channel, in Ref.~\cite{Dudek:2016cru}, the scattering matrix for $\pi \eta, K\overline{K}$ was found to feature strong channel coupling in $S$-wave, with very rapid energy dependence around the $K\overline{K}$ threshold. This was shown to be due to a resonance pole singularity lying close to the \KKbar threshold. The couplings of this resonance to the $\pi \eta$ and \KKbar channels were found to be of comparable size.

The isoscalar sector that we now turn to is notoriously more challenging for lattice QCD, owing to the need to compute completely disconnected diagrams in which all quarks and antiquarks annihilate. Using the \emph{distillation} framework, we have been able to evaluate all required contributions to correlation functions with good statistical precision. The \emph{elastic} $\pi\pi \to \pi \pi$ part of $I=0,\, S=0$ scattering (below \KKbar threshold) was presented in  Ref.~\cite{Briceno:2016mjc}, where a bound-state pole was found about 24 MeV below the $\pi\pi$ threshold. The same reference also obtained this amplitude using a lighter value of the quark masses corresponding to $m_\pi \sim 236$~MeV, and found that the bound state evolved into a broad resonance, closely resembling the experimental $\sigma$.

In calculations with $m_\pi \sim 391$~MeV, $J^P=2^+$ resonances have also been determined that are found to be narrow.  The lowest-lying $K_2^\star$ resonance in $\pi K$ is found to have a mass $m = 1576(7)$ MeV and width $\Gamma = 62(12)\, \mathrm{MeV}$ decaying into $\pi K$ only. The isovector $a_2$ was found at $m= 1506(4)$ MeV , $\Gamma = 20(3)$ MeV with approximately equal couplings to $\pi \eta$ and \KKbar. These extractions were somewhat less rigorous than those of the scalar mesons owing to the neglect of possible three-meson decays.

In this paper we will extend the work presented in Ref.~\cite{Briceno:2016mjc} at $m_\pi \sim 391\,\mathrm{MeV}$ to consider also the energy region above \KKbar threshold, and study isoscalar coupled $\pi\pi, K\overline{K}, \eta \eta$ scattering in $S$-wave and $D$-wave. We will confirm the $\sigma$ bound-state previously observed, and furthermore identify an $f_0(980)$--like state in $S$-wave, appearing close to \KKbar threshold, with strong couplings to both $\pi \pi$ and \KKbar. In the $D$-wave scattering amplitudes, we observe two clear peaks in the scattering amplitudes, which are interpreted as being due to two resonances, which resemble the experimental $f_2(1270)$ and $f_2'(1525)$, being relatively narrow, and with the lighter state dominantly coupling to $\pi\pi$ and the heavier to \KKbar.

The remainder of the paper is structured as follows: Section~\ref{Sec:spectrum} presents the finite-volume spectra determined in explicit lattice QCD calculations. Section~\ref{Sec:amps} briefly describes the technique for determining scattering amplitudes from finite-volume spectra before presenting results for elastic $\pi\pi$ scattering (\ref{Sec:elastic}), coupled $S$-wave $\pi\pi, K\overline{K}, \eta\eta$ scattering (\ref{Sec:coupled_S}) and coupled $D$-wave scattering (\ref{Sec:coupled_D}). Section~\ref{Sec:poles} examines the pole singularity content of the determined amplitudes, which is interpreted in terms of resonances in Section~\ref{Sec:interpret}. In Section~\ref{Sec:scalars}, we consider the complete flavor nonet of scalar mesons determined at $m_\pi \sim 391$~MeV, before summarizing the current calculation in Section~\ref{Sec:summary}.

%%%%% SPECTRUM %%%%%
\section{Computing the finite-volume spectrum using lattice QCD \label{Sec:spectrum}}
% !TEX root = ../f0_paper.tex

%%%%% SPECTRUM %%%%    

As in previous papers~\cite{Dudek:2009qf, Dudek:2010wm, Dudek:2010ew, Dudek:2011tt, Edwards:2011jj, Dudek:2011zz, Thomas:2011rh, Liu:2012ze, Dudek:2012ag, Dudek:2012gj, Liu:2012ze, Dudek:2012xn, Dudek:2013yja, Dudek:2014qha, Wilson:2014cna, Shultz:2015pfa, Wilson:2015dqa, Briceno:2015dca, Dudek:2016cru, Briceno:2016kkp, Briceno:2016mjc, Moir:2016srx} we make use of dynamical anisotropic Clover lattices featuring two degenerate flavors of light quark and a single heavier flavor tuned approximately to the physical strange quark mass. The quark masses are such that the pion mass is found to be close to 391 MeV. Details of the lattices, which have spatial lattice spacing $a_s \sim 0.12$~fm and a temporal spacing about three times smaller can be found in~\cite{Lin:2008pr}. In this paper, three spatial\footnote{The temporal extents are $T/a_t = 128$ except for the $20^3$ lattice, where for correlators computed in $A_1$ irreps, a lattice of temporal extent $T/a_t = 256$ is used.}
volumes are utilized: ${(L/a_s)^3 = 16^3, 20^3, 24^3}$.

For a system at rest, spectra are computed according to irreducible representations (irreps) of the cubic group, which contain \emph{subductions} of the angular momenta which characterize the infinite-volume spectrum. In systems of identical meson pairs with definite $G$-parity, only even partial waves contribute for even isospin, which subduce according to Table~II of~\cite{Dudek:2012gj}. In particular, we are interested in the irreps $A_1^+$ which contains subductions of $J^P = 0^+,4^+\ldots$, and $E^+, T_2^+$ containing $J^P=2^+,4^+\ldots$. For systems with non-zero total momentum, the symmetry is reduced to the \emph{little group} of the cubic group, which is defined by the allowed rotations of a cube which leave the total momentum invariant. In this work, we consider systems with total momentum up to $[200]$.

Stable meson masses on these lattices can be found in~\cite{Dudek:2016cru} -- particularly relevant here are pseudoscalar meson masses: ${a_t m_\pi = 0.06906(13)}$, ${a_t m_K = 0.09698(9)}$ and ${a_t m_\eta = 0.10364(19)}$. The anisotropy determined from stable meson dispersion relations is ${\xi = a_s/a_t = 3.444(6)}$.

The isoscalar sector is notoriously challenging to study within lattice QCD. In previous studies, various approximations have been made in order to make calculations practical~\cite{Alford:2000mm, *Prelovsek:2010kg, *Fu:2013ffa, *Wakayama:2014gpa, *Howarth:2015caa, *Bai:2015nea}, including the omission of disconnected diagrams or the use of only a small basis of operators. These approximations cannot in general be justified. In this paper we will compute matrices of correlation functions in a large basis of operators
\footnote{complete lists of operators used are provided in Supplemental Material.}
, including many which resemble a pair of mesons each having definite momentum, with the correlator construction achieved using the distillation framework~\cite{Peardon:2009gh}. All required Wick contractions, including those in which quarks at the source or sink annihilate, are included without further approximation. Matrices of correlation functions are analyzed variationally by solving a generalized eigenvalue problem~\cite{Michael:1985ne, Dudek:2007wv} to yield discrete spectra of states.  The operators resembling pairs of mesons with definite momentum are themselves constructed using variationally-optimized $\pi$, $K$ and $\eta$ operators~\cite{Dudek:2012gj}. The \KKbar operators are constructed with definite $G$-parity, and the $\eta\eta$ operators contain both light and strange components.  Our procedures for correlator construction and variational analysis have been described in previous papers~\cite{Peardon:2009gh,Dudek:2010wm,Thomas:2011rh,Dudek:2012gj}, and we will not repeat  details here, except to point out that we use a `weighting-shifting' correction~\cite{Dudek:2012gj} to reduce the (small) pollution due to the finite temporal extent of the lattice. For rest-frame irreps, this amounts to computing the difference between two timeslices, which also acts to remove the time-independent vacuum contribution in the $[000]A_1^+$ irrep.

\begin{figure*}
\includegraphics[width = .8\textwidth]{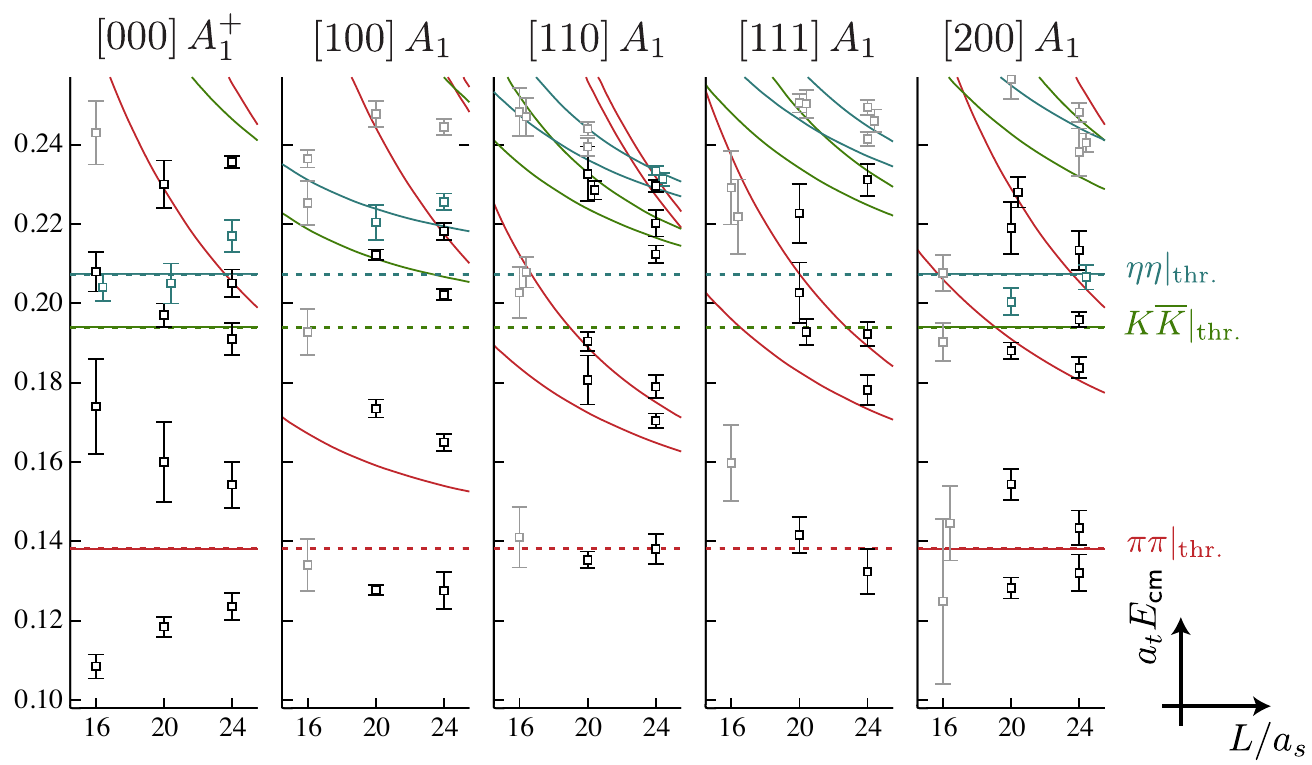}
\caption{Finite volume spectra obtained in $A_1$ irreps at rest and with four values of non-zero momentum. Dashed lines show the relevant kinematic thresholds and the solid curves the meson-meson energies in the absence of interactions: red indicates $\pi\pi$, green is \KKbar and blue is $\eta\eta$. There are no three-meson or higher thresholds in this energy region. Discrete energies determined in variation analysis of correlator matrices are plotted, and the uncertainties include systematic variation as described in the text. Energy levels colored blue are those which have large overlap with $\eta\eta$-like operators. Ghosted points are not used in the determination of coupled-channel scattering amplitudes that will be presented in Section~\ref{Sec:coupled_S}.}\label{fig:J0_spectra}
\end{figure*}

Figure~\ref{fig:J0_spectra} shows spectra determined in three lattice volumes for all $A_1$ irreps with total momentum up to $[200]$. The error bars include, as well as the statistical error due to the use of a finite sample of gauge-field configurations, also an estimate of systematic error obtained by varying the details of the variational analysis, e.g. reasonable variation of $t_0$, the extent of fitting-time windows, and the content of the variational operator basis. Significant shifts are observed with respect to non-interacting meson-meson energies shown by the curves, indicating the presence of strong scattering dynamics. The levels shown in black and blue will be used later to constrain three-channel scattering in $S$-wave. We utilise 57 energy levels densely filling an energy region from below $\pi\pi$ threshold to some way above $\eta\eta$ threshold at $a_t\, E_\mathrm{cm} \sim 0.24$.

Energy levels below $K\overline{K}$ threshold computed on these lattices previously appeared in~\cite{Briceno:2016mjc}, where they were used to determine the $\pi\pi$ elastic scattering amplitude which was found to feature a bound-state pole identified with the $\sigma$. The spectra in Figure~\ref{fig:J0_spectra} contain some small differences with respect to those presented in~\cite{Briceno:2016mjc}, typically at the level of statistical fluctuations. For this paper we have undertaken a thorough consideration of the variations seen in correlator extraction, and worked with somewhat larger operator bases in order to access the coupled-channel energy region.

%%%%%%%%%%%%%%%%%%%%%%%%%%%%%%%%%%%%%%%%%%%%%%%%%%%%%%%%%%%%%%%%%%%%%
%%%%%%%%%%%%%%%%%%%%%%%%%%%%%%%%%%%%%%%%%%%%%%%%%%%%%%%%%%%%%%%%%%%%%

\begin{figure}
\includegraphics[width = 0.95\columnwidth]{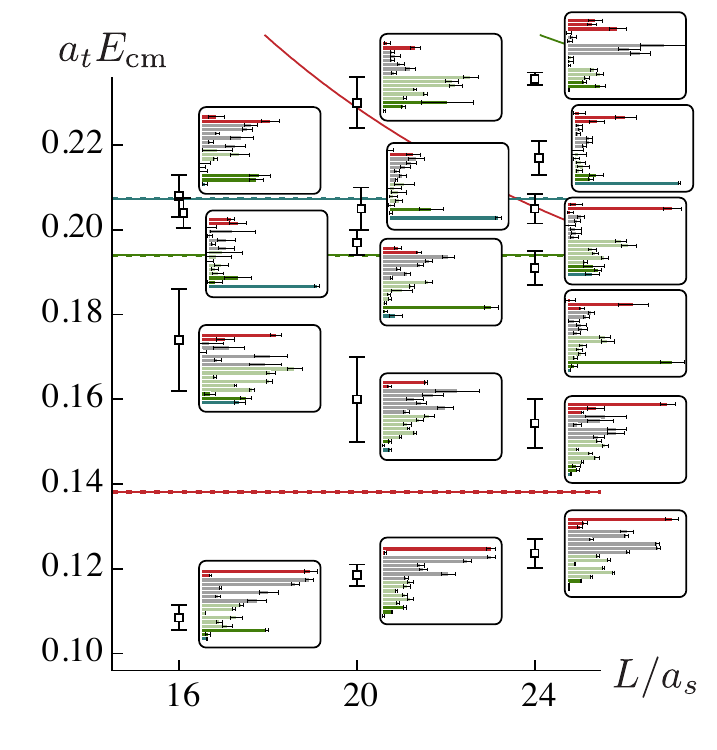}
\caption{Spectra in the $[000]A_1^+$ irrep over 3 volumes. The relative operator overlaps are shown in the histograms from top to bottom: $\pi\pi$ (red), $\bar{u}\boldsymbol{\Gamma}u + \bar{d} \boldsymbol{\Gamma} d$ (grey), $\bar{s} \boldsymbol{\Gamma}s$ (light green), $K\overline{K}$ (green), $\eta \eta$ (blue), where the normalization is defined such that for a given operator, the largest overlap over all states has unit value.}\label{fig:spec_000_A1_histo}
\end{figure}

Figure~\ref{fig:spec_000_A1_histo} shows the $[000]A_1^+$ irrep, and alongside each energy level we plot a histogram showing the relative contribution to each state from each operator in the variational basis. We highlight 5 distinct types of operator: \mbox{$\pi\pi$ (red)}, ${\bar{u}\boldsymbol{\Gamma}u + \bar{d} \boldsymbol{\Gamma} d}$ (grey), $\bar{s} \boldsymbol{\Gamma}s$ (light green), \mbox{$K\overline{K}$ (green)}, and $\eta \eta$ (blue), and the spectrum is seen to not be diagonal in this basis. A level below $\pi\pi$ is seen on all three volumes dominated by overlap with $\pi\pi$ and $\bar{u}\boldsymbol{\Gamma}u + \bar{d} \boldsymbol{\Gamma} d$ operators, which is connected to the $\sigma$ meson that appears as a shallow bound state on these lattices~\cite{Briceno:2016mjc}~\footnote{The significant volume dependence of this state suggests that it is physically large, hinting at a molecular type nature. In Sec.~\ref{Sec:scalars} we use the Weinberg criterion to quantify this speculation.}. On all three volumes we see there is a state coincident with $\eta\eta$ threshold that has dominant overlap with an operator resembling $\eta[000]\eta[000]$. The statistically negligible shift with respect to the non-interacting level, and relatively small mixing between the $\eta\eta$ operator and the other operators may be a hint that the $\eta \eta$ channel is not as strongly interacting or as strongly coupled as $\pi\pi$ and $K\overline{K}$. While operator overlaps are useful for building intuition, they are not a rigorous tool, so we reserve further comment until after the scattering amplitudes have been extracted and analyzed in Section \ref{Sec:coupled_S}.

%%%%%%%%%%%%%%%%%%%%%%%%%%%%%%%%%%%%%%%%%%%%%%%%%%%%%%%%%%%%%%%%%%%%%
%%%%%%%%%%%%%%%%%%%%%%%%%%%%%%%%%%%%%%%%%%%%%%%%%%%%%%%%%%%%%%%%%%%%%

\begin{figure}
\hspace*{-3mm}
\includegraphics[width = 1.1\columnwidth]{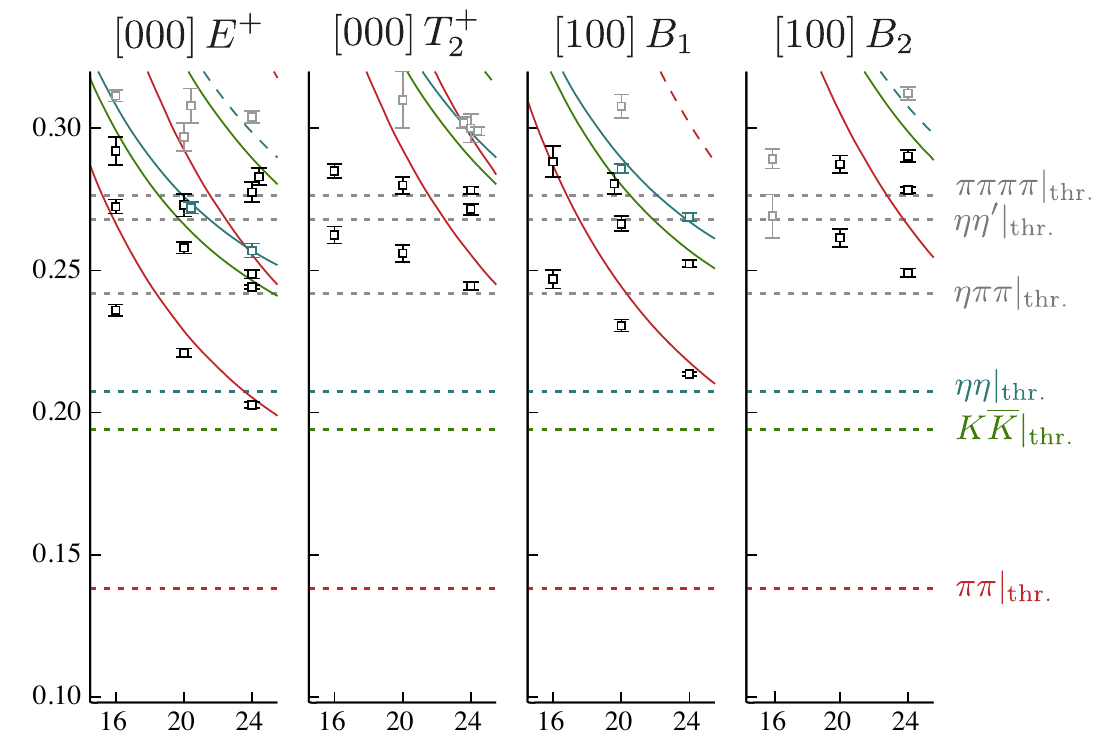}
\caption{Finite-volume spectra in irreps where $J^P=2^+$ is the lowest subduced partial wave. Dashed lines show thresholds, as in Fig.~\ref{fig:J0_spectra}, also showing the lowest neglected two--, three-- and four--body channels, $\eta \eta'$, $\eta \pi \pi$, and $\pi\pi\pi\pi$ in grey. Points highlighted in blue are dominated by overlap with an $\eta\eta$-like operator, and ghosted points are not used in the amplitude extractions in Section~\ref{Sec:coupled_D}.} \label{fig:J2_spectra}
\end{figure}

\begin{figure}
\includegraphics[width = 0.8\columnwidth]{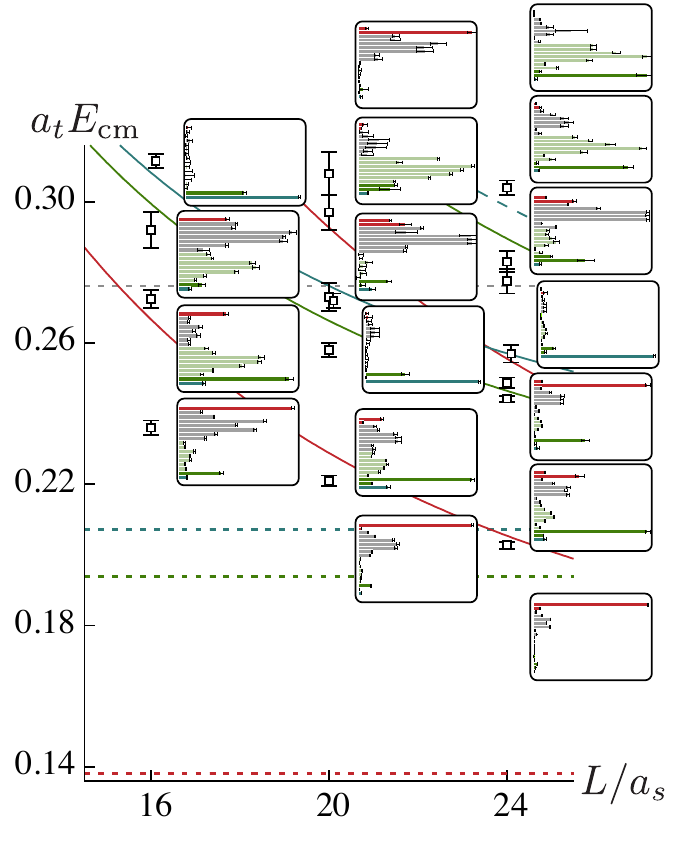}
\includegraphics[width = 0.8\columnwidth]{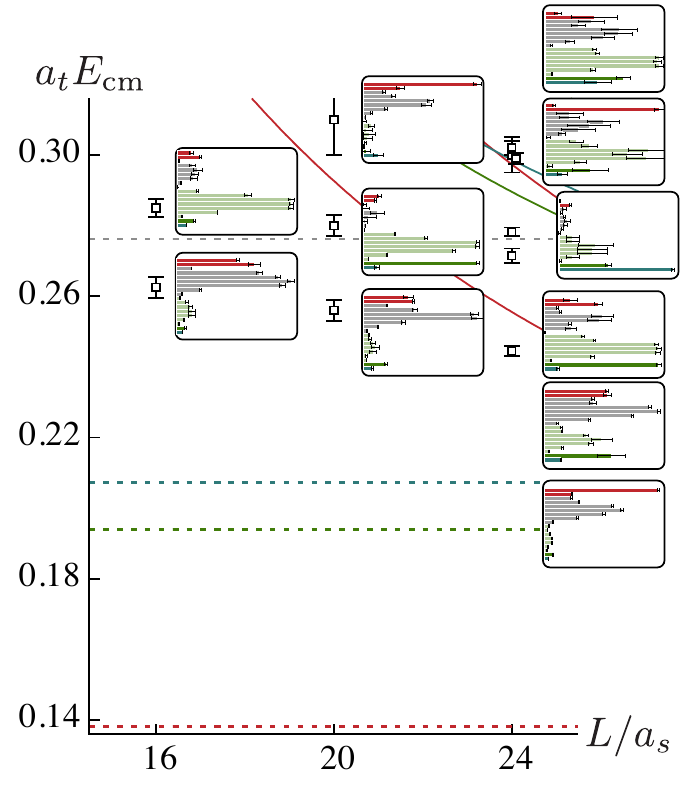}
\caption{Spectra in the $E^+$ and $T_2^+$ irreps. The histogram coloring is consistent with Fig.~\ref{fig:spec_000_A1_histo}.}\label{fig:spec_000_J2_histo}
\end{figure}

Figure~\ref{fig:J2_spectra} shows the spectra determined in three lattice volumes for irreps which feature $J^P=2^+$ scattering as the lowest angular momentum. We extract 34 energy levels shown in black and blue that are to be used in the extraction of scattering amplitudes in Section~\ref{Sec:coupled_D}. Figure~\ref{fig:spec_000_J2_histo} shows the spectrum and associated histograms for the $E^+$ and $T_2^+$ irreps, where the pattern of operator overlaps is seen to be quite different to that in $A_1^+$, with notably less `mixing' between the light and strange sectors. Statistically significant shifts from the non-interacting energies are observed, and there are more levels than would be expected based on counting the non-interacting energies on each volume in this energy region, hinting that narrow states could be present. The histograms go further, suggesting that there are likely to be two resonances, one dominated by light quarks, and another dominated by strange quarks. This is perhaps clearest in $[000]\, T_2^+$ for $L/a_s=16$ where there are no nearby non-interacting levels, but we still see two levels, one near $a_t E_\mathrm{cm} = 0.26$ dominantly overlapping with $\bar{u}\boldsymbol{\Gamma}u + \bar{d} \boldsymbol{\Gamma} d$ constructions and a second near $a_t E_\mathrm{cm} = 0.28$ dominantly overlapping with $\bar{s} \boldsymbol{\Gamma}s$ constructions. Again, we reserve further comment until after the amplitudes have been extracted in Section~\ref{Sec:coupled_D}.

%%%%%%%%%%%%%%%%%%%%%%%%%%%%%%%%%%%%%%%%%%%%%%%%%%%%%%%%%%%%%%%%%%%%%
%%%%%%%%%%%%%%%%%%%%%%%%%%%%%%%%%%%%%%%%%%%%%%%%%%%%%%%%%%%%%%%%%%%%%

%%%%%%%%%%%%%%%%%%%%%%%%%%%%%%%%%%%%%%%%%%%%%%%%%%%%%%%%%%%%%%%%%%%%%
%%%%%%%%%%%%%%%%%%%%%%%%%%%%%%%%%%%%%%%%%%%%%%%%%%%%%%%%%%%%%%%%%%%%%

We now turn to extracting the scattering amplitudes from these spectra, beginning with the $S$--wave elastic region before proceeding to investigate coupled-channel $S$-- and $D$--wave amplitudes.

%\clearpage
%%%%% AMPLITUDES %%%%%
\section{Determining coupled-channel scattering amplitudes \label{Sec:amps}}
% !TEX root = ../f0_paper.tex

%%%%% AMPS INTRO %%%%    

It has been shown that the discrete spectrum of states in a periodic finite spatial volume is determined by the energy-dependence of infinite-volume hadron-hadron scattering amplitudes~\cite{Luscher:1986pf, Luscher:1990ux, Luscher:1991cf, Rummukainen:1995vs, He:2005ey, Christ:2005gi, Kim:2005gf, Guo:2012hv, Hansen:2012tf, Briceno:2012yi, Briceno:2014oea}. The particular form of the relationship relevant to coupled-channel pseudoscalar-pseudoscalar scattering described by a $t$-matrix, $\mathbf{t}(E_\mathrm{cm})$, can be compactly expressed as~\cite{Guo:2012hv}
\begin{equation}
\det \left[ \mathbf{1} + i \boldsymbol{\rho} \cdot \mathbf{t} \cdot ( \mathbf{1} + i \boldsymbol{\mathcal{M}})\right] = 0,
\label{eq:qc}
\end{equation}
with $\boldsymbol{\rho}(E_\mathrm{cm})$ a diagonal matrix of phase-space factors, $\rho_{ij} = \delta_{ij} \frac{2 k_i}{E_\mathrm{cm}}$, and $\boldsymbol{\mathcal{M}}(E_\mathrm{cm}, L)$ a matrix of known functions which encode the `kinematics' of a cubic finite-volume. The discrete spectrum in an $L\times L \times L$ box is given by all $E_\mathrm{cm}$ for which this equation is solved. The $t$-matrix is in general not diagonal in the space of open channels, but since it is an infinite-volume quantity, it is diagonal in angular momentum. On the other hand, the finite-volume matrix $\boldsymbol{\mathcal{M}}$ is diagonal in channel space, but it is in general not diagonal in angular momentum.

This finite-volume formalism was recently reviewed in some detail in Ref.~\cite{Briceno:2017max}, and more description of our implementation (using the same notation as we will use in this paper) is given in Ref.~\cite{Dudek:2016cru}. Our approach is to parameterize the energy-dependence of the $t$-matrix, solving Eq.~\ref{eq:qc} for the discrete spectrum with a given choice of parameter values. This spectrum is then compared to the lattice spectra shown in e.g. Figure~\ref{fig:J0_spectra} in a correlated $\chi^2$ function. Minimizing the $\chi^2$ by varying parameter values we obtain a best estimate for the $t$-matrix. Subtleties associated with  broken rotational symmetry due to the cubic boundary, both at rest and in-flight, expressed mathematically by the \emph{subduction} of partial waves into \emph{irreps} of the relevant symmetry group, are discussed in Refs.~\cite{Dudek:2016cru} and \cite{Thomas:2011rh}.

One approach to ensure that the $t$-matrix in \mbox{partial-wave} $\ell$ satisfies multichannel unitarity (required in order for Eq.~\ref{eq:qc} to have solutions) is to express it in terms of a $K$-matrix,
\begin{equation}
(t^{-1})_{ij} = \frac{1}{(2k_i)^\ell} (K^{-1})_{ij} \frac{1}{(2k_j)^\ell} + I_{ij},
\label{eq:K}
\end{equation}
where the factors $(2 k_i)^{-\ell}$ provide the required kinematic behavior at threshold for the $\ell$-wave. The elements $K_{ij}(E_\mathrm{cm})$ form a symmetric matrix that is real for real values of $E_\mathrm{cm}$, and the elements $I_{ij}$ form a diagonal matrix whose imaginary part is fixed above threshold by unitarity to be $- \rho_i$. An option for the real part of $\mathbf{I}$ which ensures the amplitude behaves sensibly below threshold and for complex values of the energy is to use the Chew-Mandelstam phase-space -- a discussion can be found in Ref.~\cite{Wilson:2014cna}.

We will make use of a range of parameterizations for the matrix $\mathbf{K}$ in the $S$-- and $D$--waves, considering mostly the coupled three-channel problem $\pi\pi, K\overline{K}, \eta\eta$. We will later also have cause to also consider an application of the Jost functions, through which we may explicitly control the singularity content of the amplitudes, at the cost of losing guaranteed unitarity for all parameter values. The possibility of extracting information about scalar mesons from lattice QCD spectra has previously been discussed in the context of various amplitude parameterizations in Refs.~\cite{Bernard:2010fp, Doring:2011vk, Doring:2011ip, Doring:2012eu}.

  \subsection{Elastic $S$-wave $\pi\pi$ scattering \label{Sec:elastic}}
  % !TEX root = ../f0_paper.tex

%%%%% ELASTIC %%%%    

An analysis of the discrete spectrum of states in the energy region below $K\overline{K}$ in terms of elastic $\pi\pi$ scattering in $S$-wave was previously presented in~\cite{Briceno:2016mjc}. In this case, under the (verified) assumption that higher partial-waves make a negligible contribution to Eq.~\ref{eq:qc}, the problem reduces to a one-to-one mapping from the computed $E_\mathrm{cm}$ value to a value of the scattering amplitude at that energy. In this way the elastic scattering amplitude was mapped out, and a behavior compatible with a bound-state that we associated with the $\sigma$ was observed.

\begin{figure}[b]
\includegraphics[width = 0.95\columnwidth]{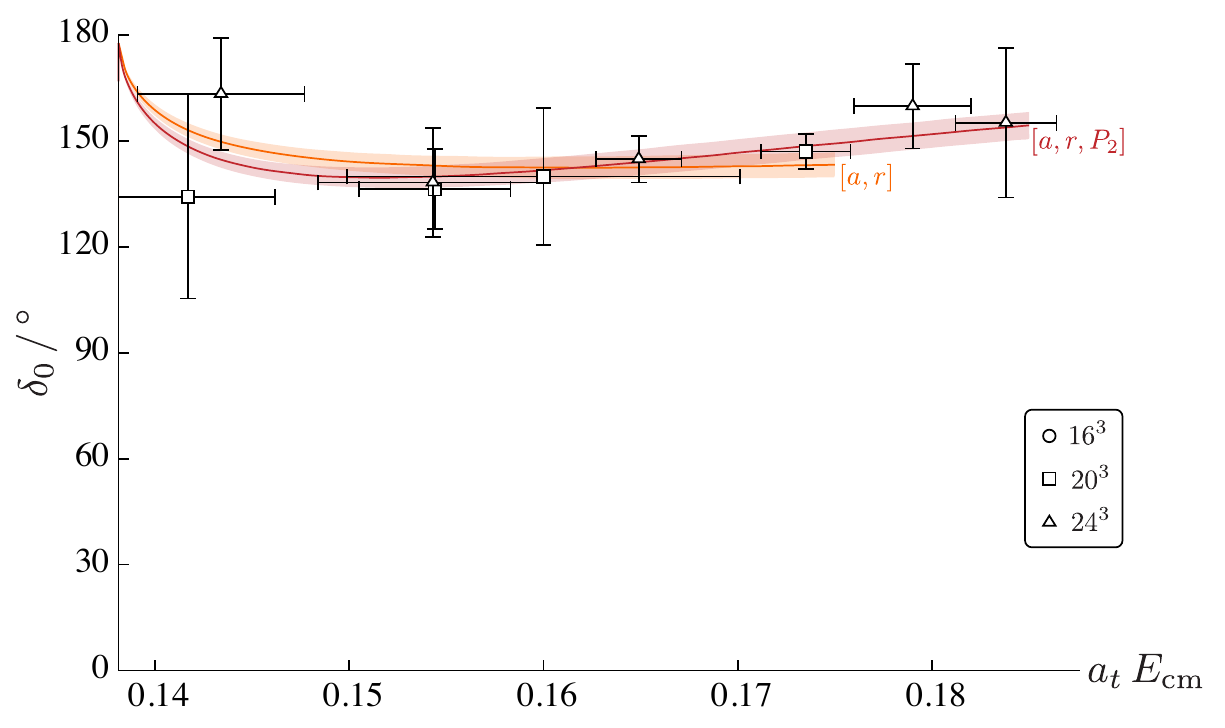}
\includegraphics[width = 0.95\columnwidth]{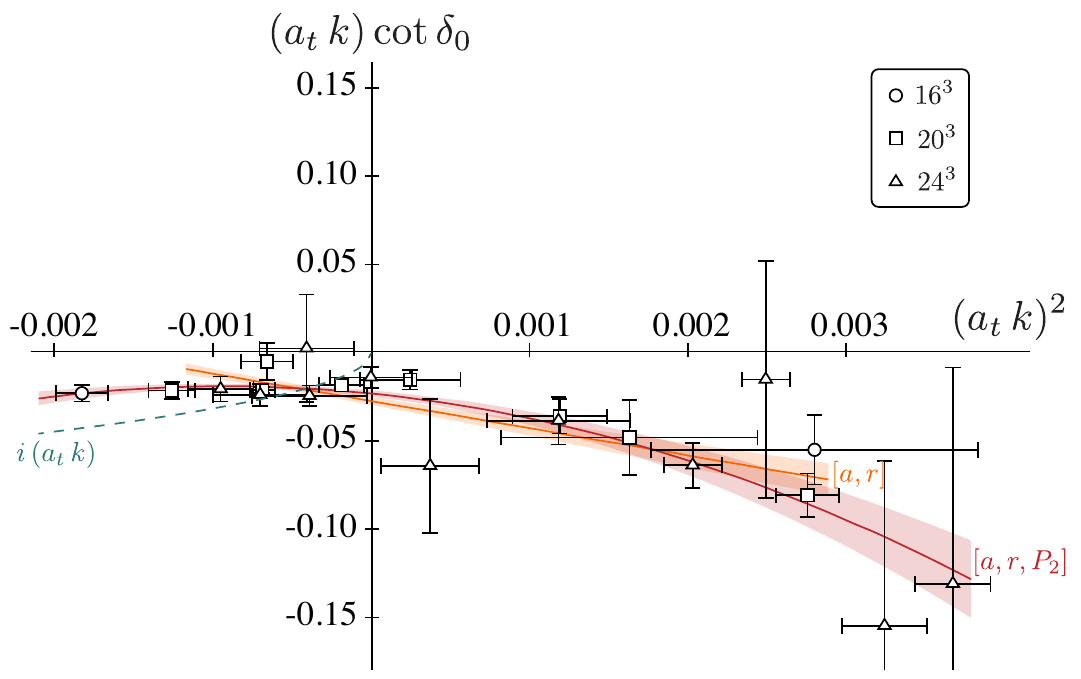}
\caption{$S$-wave $\pi\pi$ elastic scattering amplitude expressed by the phase-shift, $\delta_{\pi\pi}$ (top) and $k\,\cot \delta$ (bottom). Points with large error bars are not plotted for clarity. Curves indicate effective-range-expansion descriptions discussed in the text.}\label{fig:elastic_delta}
\end{figure}

For this paper we have considered larger matrices of correlation functions, including as well as those used in Ref.~\cite{Briceno:2016mjc}, also a significant number of $K\overline{K}$ and $\eta\eta$ operators, and the resulting variational analysis leads to spectra that differ slightly from those presented in~\cite{Briceno:2016mjc}, in particular through better estimation of a systematic error. We use these improved energy levels here to determine the elastic scattering amplitude, expressed via the phase-shift ($t_{\pi\pi, \pi\pi} = \rho^{-1}_{\pi\pi}\,  e^{i \delta_{\pi\pi}} \sin \delta_{\pi\pi}$), assuming that $\ell=2$ and higher partial waves have a negligible impact on the finite-volume spectrum\footnote{We will actually determine these $\ell=2$ amplitudes in Section~\ref{Sec:coupled_D} and find that their effect on the $A_1$ irreps at low energies is negligible.}. In Figure~\ref{fig:elastic_delta} we present discrete phase-shift points, as well as effective-range-expansion descriptions of the finite-volume spectrum below $K\overline{K}$ threshold.

%%%%%%%%%%%%%%%%%%%%%%%%%%%%%%%%%%%%%%%%%%%%%%%%%%%%%%%%%%%%%%%%%%%%%
%%%%%%%%%%%%%%%%%%%%%%%%%%%%%%%%%%%%%%%%%%%%%%%%%%%%%%%%%%%%%%%%%%%%%

Re-expressing the elastic $t$-matrix as $t = \frac{E_\mathrm{cm}}{2} \frac{1}{k \cot \delta - i k}$ makes it clear that if a graph of $k \cot \delta$ intersects the curve $i k$ at energies below threshold, where $ik = - |k|$, there will be a bound-state pole singularity. Such a crossing is clearly visible in the lower pane of Figure~\ref{fig:elastic_delta}, corresponding to a bound-state $\sigma$. We will explicitly address the singularity content of scattering amplitudes and their interpretation in terms of bound-states and resonances in Section~\ref{Sec:poles}. 

The energy dependence of the elastic amplitude can be described within the $S$--wave effective range expansion, ${k \cot \delta_0 = \frac{1}{a} + \tfrac{1}{2} r k^2 + \sum_{n=2} P_n \,k^{2n}}$. Describing those energies in the region $ 0.12 \le a_t E_\mathrm{cm} \le 0.175$, by a scattering length plus effective range form we obtain $a/a_t = -36.6 \pm 3.2$, $r/a_t = -30.9 \pm 5.2$ with a ${\chi^2/N_\mathrm{dof} = 16.3 / (17-2) = 1.09}$. A larger energy region, $0.1 \le a_t E_\mathrm{cm} \le 0.185$, can be described if we include also a $k^4$ term in the expansion, yielding $a/a_t = -43.1 \pm 3.9$, $r/a_t = -18.1 \pm 4.4$ with a ${\chi^2/N_\mathrm{dof} = 18.1 / (23-2) = 0.91}$. Both descriptions are displayed in Figure~\ref{fig:elastic_delta}. In this elastic analysis we avoid considering levels lying just below the $K\overline{K}$ threshold, as in finite-volume they are impacted by the $t_{\pi\pi \to K\overline{K}}$ and $t_{K\overline{K} \to K\overline{K}}$ elements of the $t$-matrix~\cite{Guo:2012hv, Briceno:2017max}.

  \subsection{Coupled $S$-wave $\pi\pi$, $K\overline{K}$, $\eta \eta$ scattering \label{Sec:coupled_S}}
  % !TEX root = ../f0_paper.tex

%%%%% COUPLED S-wave %%%%   
A na\"ive examination of the energy levels presented in Figure~\ref{fig:J0_spectra}, where an `extra' level is present below $\pi\pi$ threshold, strongly suggests the existence of the bound-state $\sigma$ discussed in the previous section, but such simple analysis does not obviously indicate any narrow resonances at higher energy -- while significant shifts of energy levels away from non-interacting energies are observed, there are not clearly any `extra' levels that one might associate with a nearly-stable state. 

We will proceed by attempting to describe 57 energy levels lying below $a_t E_\mathrm{cm} = 0.24$ shown in black and blue in Figure~\ref{fig:J0_spectra} using a range of three-channel ($\pi\pi$, $K\overline{K}$, $\eta\eta$) $S$-wave amplitude forms in Eq.~\ref{eq:qc}. In this section we will make use of $K$-matrix parameterizations as described previously. 
%Later, when we consider the resonance pole content of the scattering system, we will also explore Jost-style amplitudes.

An illustrative example of a successful description of the spectrum is provided by a $K$-matrix in which the matrix inverse of $K$ is parameterized as
\begin{equation}
\mathbf{K}^{-1}(s) = \begin{pmatrix} 
a + b \,s   &   c + d \, s   &  e \\ 
c + d \, s   &   f         &  g \\
e         &   g         &  h
\end{pmatrix},
\label{eq:ref_amp}
\end{equation}
where the row (and column) channel ordering is $\pi\pi, \, K\overline{K}, \, \eta \eta$. There are eight free parameters, namely the constants $a\ldots h$, and the Chew-Mandelstam phase-space used for $\mathbf{I}$ is subtracted for each channel at the relevant threshold. The best description of the spectra with this amplitude
\footnote{The fitted parameter values may be found in Supplemental Material.}
, shown in Figure~\ref{fig:model_spec}, has ${\chi^2/N_\mathrm{dof} = 44.0 / (57 -8) = 0.90}$.

%%%%%%%%%%%%%%%%%%%%%%%%%%%%%%%%%%%%%%%%%%%%%%%%%%%%%%%%%%%%%%%%%%%%%
\begin{figure}[b]
\hspace*{-6mm}\includegraphics[width = 1.05\columnwidth]{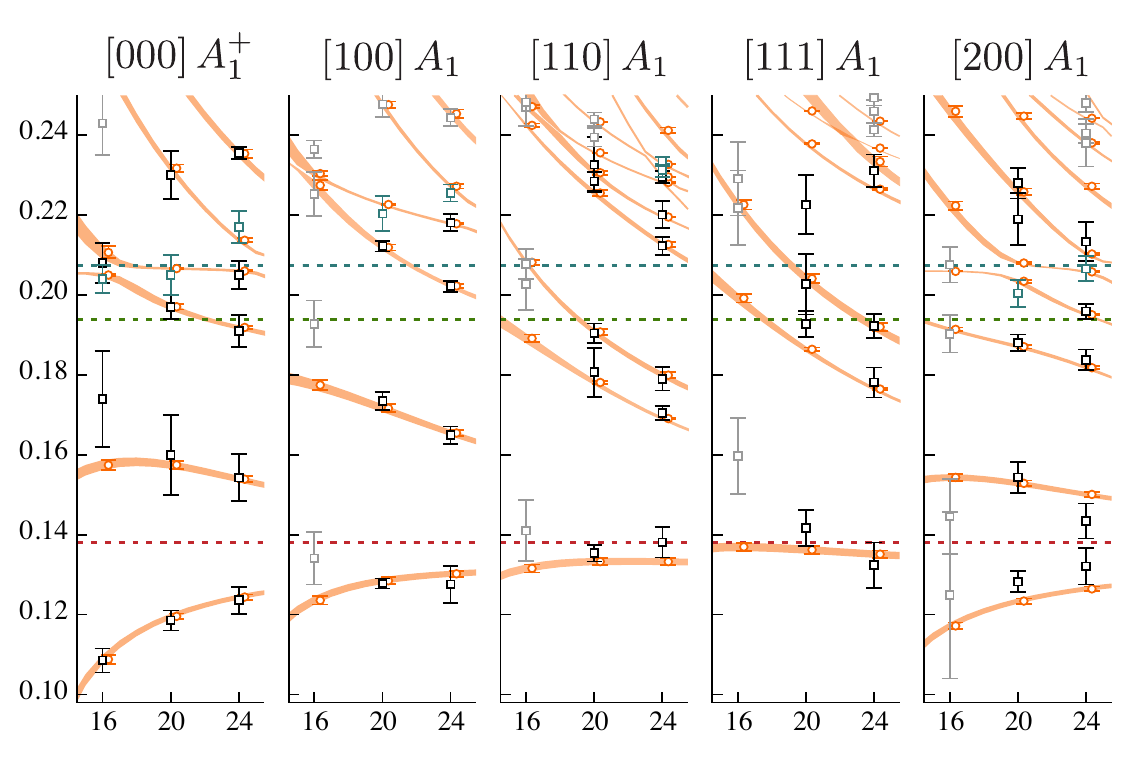}
\caption{$A_1$ irrep spectra as described by the amplitude given in Eq.~\ref{eq:ref_amp}. Black/blue points show the lattice QCD spectrum of Figure~\ref{fig:J0_spectra} and orange curves and points (displaced slightly for clarity) the result of the fitted amplitude.}\label{fig:model_spec}
\end{figure}
%%%%%%%%%%%%%%%%%%%%%%%%%%%%%%%%%%%%%%%%%%%%%%%%%%%%%%%%%%%%%%%%%%%%%

This is clearly a very successful description of the lattice QCD energy levels -- the fact that this reduced $\chi^2$ (and that of many other successful descriptions) is slightly below 1 may indicate that we have been a little too conservative in our estimation of systematic error on the energy levels. We note that the fit to 57 energy levels shown in bold also provides a quite reasonable description of levels at slightly higher energies, and on the ${L/a_s=16}$ lattice in-flight (shown ghosted), that we conservatively did not include in the fit.

\begin{figure*}
\includegraphics[width = 0.75\textwidth]{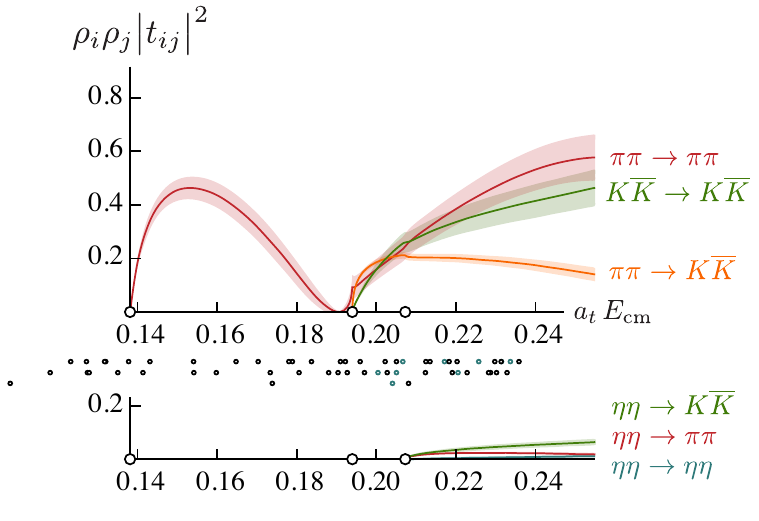}
\caption{The $S$-wave scattering amplitude expressed as e.g. $\rho_{\pi\pi}^2 \big| t_{\pi\pi, \pi\pi} \big|^2$(red), for the amplitude given by Eq.~\ref{eq:ref_amp}. The bands indicate the uncertainty obtained by propagating through the calculation the correlated uncertainties on the energy levels presented in Figure~\ref{fig:J0_spectra}. The thresholds for $\pi\pi$, $K\overline{K}$ and $\eta\eta$ are indicated by the circles on the energy axis, and the discrete energy levels used to constrain the amplitude are displayed by the small dots appearing under the energy axis, with blue dots indicating those levels having large overlap onto $\eta\eta$-like operators.}\label{fig:k_inv_poly_llcccc_rho_t_sq}
\end{figure*}

The energy dependence of the resulting amplitude is shown in Figure~\ref{fig:k_inv_poly_llcccc_rho_t_sq}, where what is plotted is a quantity that is proportional to the cross-section for various scattering processes. Above a broad bump in $\pi\pi \to \pi\pi$ at low energies caused by the tail of the $\sigma$ bound-state, is a sharp dip just below the opening of the $K\overline{K}$ threshold. Along with a slight cusp in $\pi\pi \to \pi\pi$ at $K\overline{K}$ threshold, we observe a rapid turn on of the $K\overline{K}$ channel. There is clearly very little activity in $\eta \eta$, and only tiny kinks in the $\pi\pi$ and $K\overline{K}$ channels at $\eta\eta$ threshold. The small circles shown alongside the energy axis indicate the 57 energy levels used to constrain the amplitude -- we note that they densely span the entire energy region, overconstraining the energy dependence of the $t$-matrix. The statistical uncertainties on the $\pi$, $K$ and $\eta$ masses prove to have a negligible effect on the amplitude determination.

In passing we note that the rest-frame spectrum in the three volumes considered alone would have provided us with only 15 energy levels. Of these only 9 are in the coupled-channel region and only three of these have large $\eta\eta$ overlap. Such limited information would have provided minimal constraint on our parameterization and consequently would have given us little confidence in our final result. 

%%%%%%%%%%%%%%%%%%%%%%%%%%%%%%%%%%%%%%%%%%%%%%%%%%%%%%%%%%%%%%%%%%%%%
%%%%%%%%%%%%%%%%%%%%%%%%%%%%%%%%%%%%%%%%%%%%%%%%%%%%%%%%%%%%%%%%%%%%%

\begin{figure}
\includegraphics[width = 1.05\columnwidth]{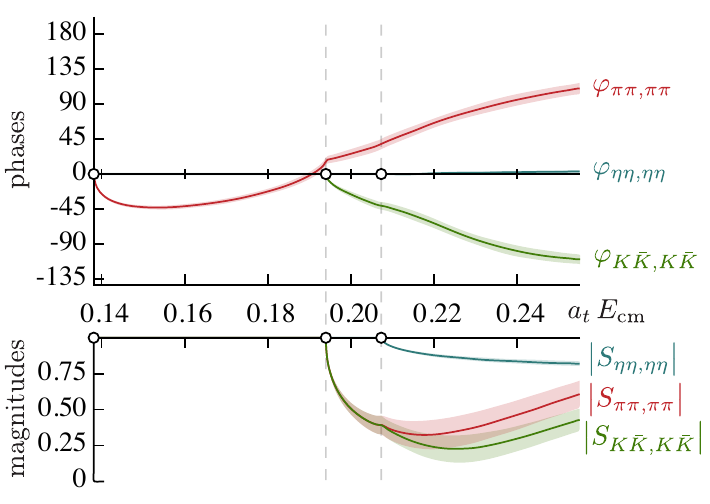}
\caption{Phases and magnitudes of diagonal elements of the $S$-matrix, $S_{ii}(E_\mathrm{cm}) = \big|S_{ii}(E_\mathrm{cm})\big| \,e^{2 i \varphi_{ii}(E_\mathrm{cm})}$, for the amplitude given by Eq.~\ref{eq:ref_amp}. }\label{fig:k_inv_poly_llcccc_diagS}
\end{figure}

Another way to present the energy dependence of the scattering amplitudes is by plotting the magnitudes and phases of the diagonal elements of the $S$-matrix
\footnote{ $\mathbf{S} = \mathbf{1} + 2 i \sqrt{\boldsymbol{\rho}} \cdot \mathbf{t} \cdot \! \sqrt{\boldsymbol{\rho}}$ }
, $S_{ii}(E_\mathrm{cm}) = \big|S_{ii}(E_\mathrm{cm})\big| \,e^{2 i \varphi_{ii}(E_\mathrm{cm})}$. The two-channel unitarity constraint is such that between the $K\overline{K}$ and $\eta\eta$ thresholds, $\big| S_{\pi\pi,\pi\pi} \big| = \big| S_{K\overline{K},K\overline{K}} \big|$ and this magnitude can be associated with what is usually called the (in)elasticity, $\eta$, while the phases $\varphi_{\pi\pi, \pi\pi},\, \varphi_{K\overline{K}, K\overline{K}}$ are identified with the channel phase-shifts, $\delta_{\pi\pi}, \delta_{K\overline{K}}$. Above the $\eta\eta$ threshold, three-channel unitarity comes into play, and such simple identifications can no longer be made, although the relatively weak coupling to the $\eta\eta$ channel suggests that a decoupled ``$2+1$'' channel (coupled $\pi\pi, K\overline{K}$ plus decoupled $\eta\eta$) description might be a reasonable approximation. Figure~\ref{fig:k_inv_poly_llcccc_diagS} shows these phases and magnitudes.

%%%%%%%%%%%%%%%%%%%%%%%%%%%%%%%%%%%%%%%%%%%%%%%%%%%%%%%%%%%%%%%%%%%%%
%%%%%%%%%%%%%%%%%%%%%%%%%%%%%%%%%%%%%%%%%%%%%%%%%%%%%%%%%%%%%%%%%%%%%

%%%%%%%%%%%%%%%%%%%%%%%%%%%%%%%%%%%%%%%%%%%%%%%%%%%%%%%%%%%%%%%%%%%%%
\begin{figure}
\includegraphics[width = \columnwidth]{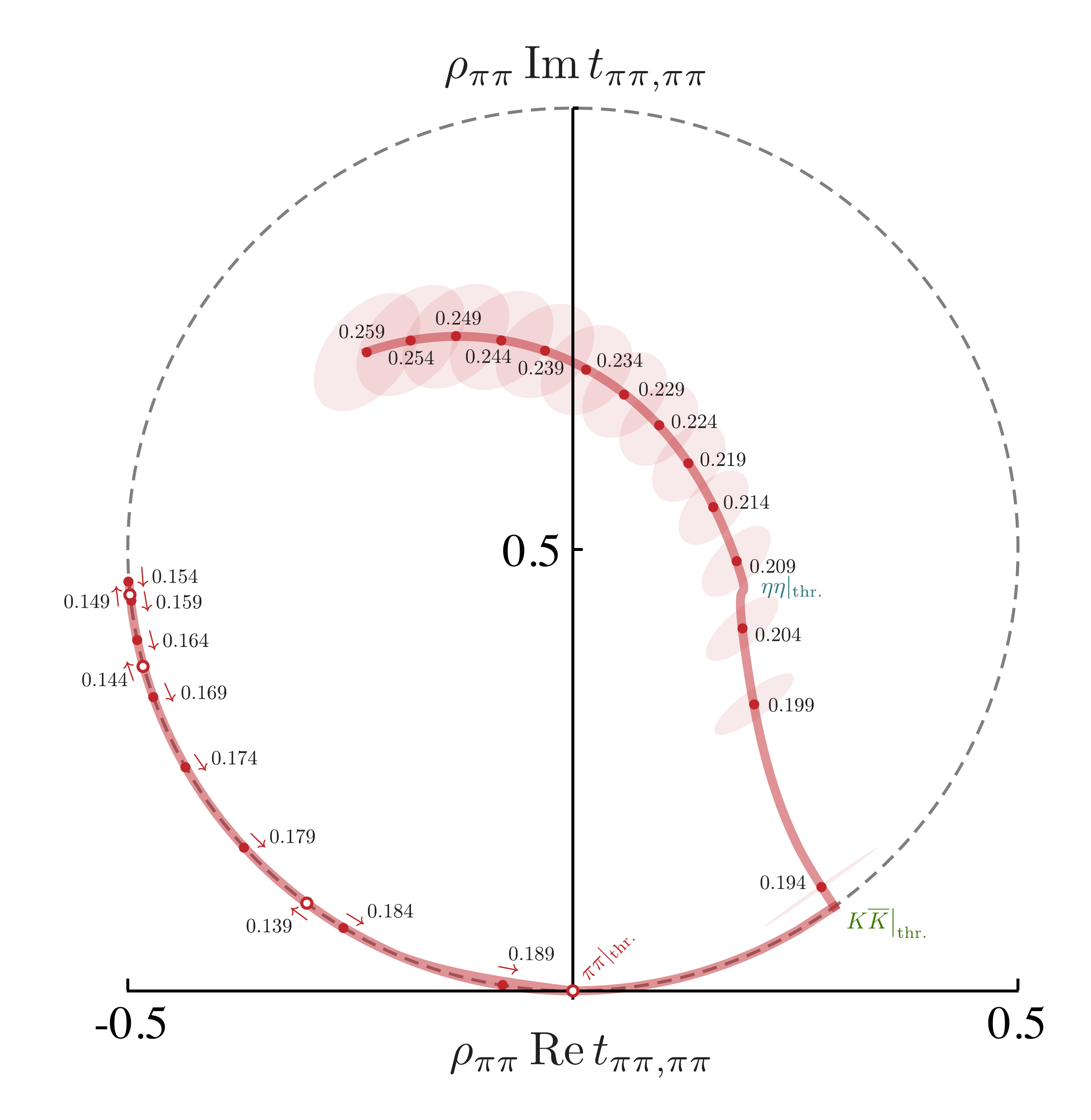}
\caption{Argand diagram representation of the $t_{\pi\pi, \pi\pi}$ element of the amplitude given by Eq.~\ref{eq:ref_amp}. Points are spaced evenly in energy with $a_t \Delta E_\mathrm{cm} = 0.005$. The amplitude initially moves clockwise from $\pi\pi$ threshold (open circles) before doubling back upon itself (closed circles).}\label{fig:k_inv_poly_llcccc_argand_pi}
\end{figure}
%%%%%%%%%%%%%%%%%%%%%%%%%%%%%%%%%%%%%%%%%%%%%%%%%%%%%%%%%%%%%%%%%%%%%

A time-honored illustration of the `elastic' part of the scattering matrix is provided by the Argand diagram, which is shown for $\pi\pi \to \pi\pi$ in Figure~\ref{fig:k_inv_poly_llcccc_argand_pi}. Starting at $\pi\pi$ threshold, the amplitude initially moves rapidly clockwise along the unitarity circle under the influence of the $\sigma$ bound-state, before doubling back and more slowly passing though the point $t_{\pi\pi, \pi\pi} = 0$ corresponding to the dip in Figure~\ref{fig:k_inv_poly_llcccc_rho_t_sq}. Shortly after this, upon reaching the $K\overline{K}$ threshold, the amplitude moves inside the unitarity circle as probability is lost from $\pi\pi$ into the $K\overline{K}$ channel. The opening of the $\eta\eta$ channel is marked by only a small kink in the curve.

Examining Figures~\ref{fig:k_inv_poly_llcccc_rho_t_sq}, \ref{fig:k_inv_poly_llcccc_diagS}, \ref{fig:k_inv_poly_llcccc_argand_pi}, it is clear that we are dealing with a rather non-trivial scattering system whose resonant content cannot immediately be inferred. Certainly the amplitude looks nothing like the traditional view of a simple resonance appearing as a clearly defined bump on a slowly varying background. However, it is important to note that by fully respecting unitarity, significant constraints have been placed on the possible energy dependence of the amplitude, particularly in the elastic region -- this, coupled with the presence of the $\sigma$ bound-state, could cause the appearance of a resonance to be significantly distorted with respect to our na\"ive intuition.

A rigorous definition of the resonant content of a scattering system will come from examining the $t$-matrix at complex values of $s$ where pole singularities will appear if the system contains resonances. We will explore the pole content in a later section, but at this stage it is worth noting that the sharp dip in $\pi\pi$ and rapid turn on of $K\overline{K}$ intensity suggests there may be a singularity in the vicinity of the $K\overline{K}$ threshold. We also note in passing that although the $\sigma$ at this quark mass appears as a bound-state rather than the broad resonance seen in experiment, the energy dependence observed in Figure~\ref{fig:k_inv_poly_llcccc_rho_t_sq} is not dissimilar to that extracted from pion beam experiments~\cite{Protopopescu:1973sh, Hyams:1973zf, Grayer:1974cr, Kaminski:1996da} as shown in, for example, Fig.~1 of Ref.~\cite{Pennington:1988bp}.

%%%%%%%%%%%%%%%%%%%%%%%%%%%%%%%%%%%%%%%%%%%%%%%%%%%%%%%%%%%%%%%%%%%%%%%%%%%%%%
%%%%%%%%%%%%%%%%%%%%%%%%%%%%%%%%%%%%%%%%%%%%%%%%%%%%%%%%%%%%%%%%%%%%%%%%%%%%%%
%%%%%%%%%%%%%%%%%%%%%%%%%%%%%%%%%%%%%%%%%%%%%%%%%%%%%%%%%%%%%%%%%%%%%%%%%%%%%%

\subsubsection{Varying the amplitude parameterization}

In the previous section we discussed one particular \mbox{$K$-matrix} amplitude parameterization that successfully described the lattice QCD spectra of Figure~\ref{fig:J0_spectra}. We have explored a wide variety of parameterizations, and in this section we will report on the variation observed in the resulting amplitudes. For clarity of presentation we have selected an illustrative set of 20 parameterizations (including the one presented in the previous section), all of which have $\chi^2/N_\mathrm{dof}$ below 1.05, and which show no excessive parameter correlation\footnote{we also reject any amplitude which we find to feature a nearby off-axis pole singularity on the physical sheet -- such behavior is acausal and reflects the lack of explicit analyticity constraints on our $K$-matrix amplitudes.}. Many other parameterizations were found to successfully describe the lattice QCD spectra with behavior compatible with one or more of the 20 shown here\footnote{Details of the 20 fits presented here are provided in Supplemental Material.}. 

Figure~\ref{fig:J0_amps} shows the variation in central values for 19 parameterizations along with the reference amplitude described in the previous section, and what is clear is that the elastic region behavior, including the position of the dip, shows very little variation, and neither does the behavior of amplitudes in the region between the $K\overline{K}$ threshold and the $\eta \eta$ threshold. The bulk of variation under change of amplitude form lies at and above the $\eta \eta$ threshold, and is ultimately associated with how strongly the fit allows the $\eta \eta$ channel to couple into the strongly coupled $\pi\pi$, $K\overline{K}$ system. The variation is, however, only modest, with the qualitative behavior of the amplitudes being the same for all these successful fits, and in particular the rapid energy dependence around the $K\overline{K}$ threshold is rather well pinned down.

%%%%%%%%%%%%%%%%%%%%%%%%%%%%%%%%%%%%%%%%%%%%%%%%%%%%%%%%%%%%%%%%%%%%%
\begin{figure}
\hspace*{-4mm}\includegraphics[width = 1.15\columnwidth]{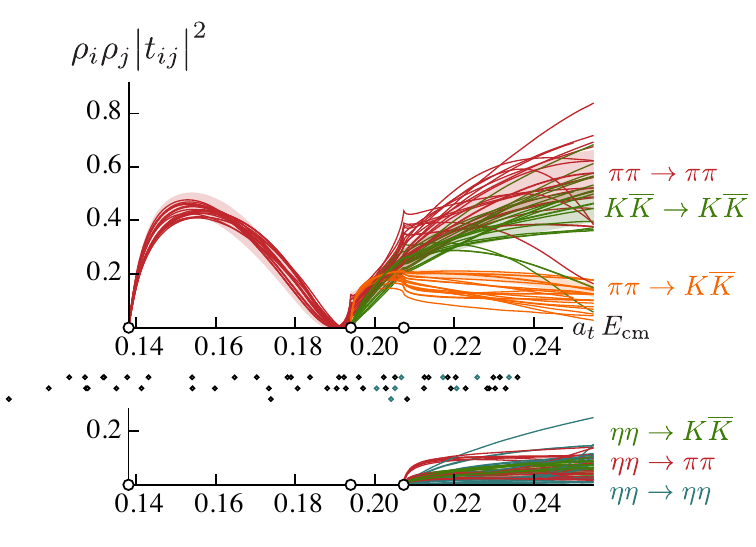}
\caption{Thick lines and bands show the amplitude previously plotted in Figure~\ref{fig:k_inv_poly_llcccc_rho_t_sq}. Thinner lines show the central value of amplitudes from 19 further parameterizations which successfully describe the lattice QCD spectra.}\label{fig:J0_amps}
\end{figure}
%%%%%%%%%%%%%%%%%%%%%%%%%%%%%%%%%%%%%%%%%%%%%%%%%%%%%%%%%%%%%%%%%%%%%

\vspace{5mm}
The in-flight $A_1$ irreps used to constrain the amplitudes do receive contributions from ${\ell=2}$ amplitudes, although we expect these to be significantly suppressed at these low energies by the centrifugal barrier. We have explicitly estimated their effect by including in our fits the $\ell=2$ amplitudes which successfully describe the spectra presented in Figure~\ref{fig:J2_spectra}, to be discussed in the next section. When these amplitudes were included in Eq.~\ref{eq:qc} no significant change in the \mbox{$S$-wave} amplitudes was observed and we conclude that \mbox{$D$-wave} amplitudes play a negligible role in determining the spectra presented in Figure~\ref{fig:J0_spectra}.

\vspace{5mm}

One of the amplitudes included in our illustrative set of 20 includes an Adler zero, implemented by multiplying a parameterization of $\mathbf{K}(s)$ by a factor $(s - s_A)$ with the position of the zero set at the value suggested by leading-order chiral perturbation theory, $s_A = \frac{1}{2} m_\pi^2$. The resulting fitted amplitude does not show any noteworthy differences with respect to the others considered. We explored the dependence on the position of the zero, by varying $s_A$ in the region from $-5 m_\pi^2$ up to $\frac{3}{2} m_\pi^2$, allowing the other parameters in the amplitude to float freely. We observed that the $\chi^2$ was largely insensitive to the position of the zero, with a very slight preference for large negative values (where its presence becomes of decreasing importance). The appearance of the amplitudes hardly varies under change in $s_A$, and the pole singularity content of the amplitudes (to be discussed in Section~\ref{Sec:poles}) is similarly insensitive. We conclude that an Adler zero is not an important feature of the $S$-wave $t$-matrix at \mbox{$m_\pi \sim 391$ MeV}.

% ref_amp_k_inv_poly_llcccc			chisq = 0.90
% amp1_k_inv_poly_lccccc			chisq = 0.92
% amp2_k_inv_poly_lcclcc			chisq = 0.93
% amp3_k_inv_poly_qccccc			chisq = 0.93
% amp4_k_inv_poly_lccccl			chisq = 0.92
% amp5_k_inv_poly_clcccc_noCM		chisq = 0.97
% amp6_k_inv_poly_ccclcc_noCM		chisq = 1.00
% amp7_k_inv_poly_llcccc_noCM		chisq = 0.97
% amp8_k_inv_poly_lcclcc_noCM		chisq = 1.01
% amp9_k_lcclcc						chisq = 1.00
% amp10_k_lcclcl					chisq = 1.02
% amp11_k_qcclcc					chisq = 0.99
% amp12_k_lcclcc_noCM				chisq = 1.03
% amp13_k_llclcc_noCM				chisq = 0.96
% amp14_k_sigma_pole_cccccc			chisq = 0.92
% amp15_k_sigma_pole_lccccc			chisq = 0.92
% amp16_k_sigma_pole_ccclcc			chisq = 0.91
% amp17_k_sigma_pole_llcccc			chisq = 0.90
% amp18_k_sigma_pole_sccccc_Adler_LO chisq = 0.97
% amp19_k_sigma_pole_ssssss_noCM	chisq = 1.03

  \subsection{Coupled $D$-wave $\pi\pi$, $K\overline{K}$, $\eta \eta$ scattering \label{Sec:coupled_D}}
  % !TEX root = ../f0_paper.tex

%%%%% COUPLED D-wave %%%%   

The lattice QCD energy levels we obtained in irreps $[000]\, E^+$, $[000]\, T_2^+$, $[100]\, B_1$, and $[100]\, B_2$ have $\ell=2$ scattering as their lowest subduced contribution. We have attempted to describe the 34 energy levels shown in bold in Figure~\ref{fig:J2_spectra}, assuming three-channel scattering, $\pi\pi$, $K\overline{K}$, $\eta\eta$, and ignoring contributions from $\ell\geq4$. We note that only four levels have significant overlap with \mbox{$\eta \eta$-like} operator constructions, which suggests that we will have only limited constraint on the $\eta \eta$ sector. However, the proximity of these four levels to the corresponding non-interacting $\eta \eta$ levels suggests that the channel may well be only weakly interacting, and that solutions in which $\eta \eta$ is largely decoupled from $\pi\pi, K\overline{K}$ may be successful.

In Figure~\ref{fig:J2_spectra} we observe the opening of the $\eta \eta'$ channel within the energy region under consideration. We considered the difference in the spectrum extracted including and not including $\eta \eta'$ operators and observed no statistically significant change, beyond addition of poorly determined energy levels at higher energies. We will neglect the effect of this channel.

In Figure~\ref{fig:J2_spectra} we also see that a three-hadron channel, $\eta \pi \pi$, and a four-hadron channel, $\pi\pi\pi\pi$, open in the energy region being considered. We did not include any operators resembling these channels when we computed the correlation matrices. A complete formalism for relating finite-volume spectra to amplitudes including \mbox{three-hadron} and higher multiplicity scattering does not yet exist, but progress in that direction is being made~\cite{Briceno:2017tce, Hammer:2017kms, Hansen:2014eka, Hansen:2015zga, Briceno:2012rv, Polejaeva:2012ut}. We do have reason to believe that these channels will not have a significant impact at the energies we are considering. Experimentally, in virtually all three-hadron and higher multiplicity processes, the final state is dominated by the parts of the phase-space where two \mbox{(or more)} hadrons resonate through an \emph{isobar}. In this case, the lowest-lying contributing isobar systems for $\eta \pi \pi$ with $J=2$ would be $a_2 \pi$ and $f_2 \pi$, and in~\cite{Dudek:2016cru}, the $a_2$ decaying to $\pi \eta$ was found with a mass $a_t m \sim 0.26$. Later in this paper we will find the lightest $f_2$ resonance at a similar mass. Thus we might estimate the energies at which we expect $\eta \pi \pi$ to become a considerable influence to be above $a_t E_\mathrm{cm} \sim 0.33$. A possible contribution from the isobar process $\rho\rho$ in $\pi\pi\pi\pi$ would be expected above $a_t E_\mathrm{cm} \sim 0.30$, and this in part motivates our decision not to consider in this first analysis any levels above $a_t E_\mathrm{cm} \sim 0.30$.

As was discussed in Section~\ref{Sec:spectrum}, spectra and overlaps suggest there may be two narrow enhancements near ${a_t E_\mathrm{cm} = 0.26, 0.28}$, and an efficient way to allow for two narrow resonances within a multichannel $K$-matrix is to use a parameterization which includes two real poles. When `dressed' by the phase-space, real poles in the \mbox{$K$-matrix} with relatively small residues can give rise to an amplitude which resembles a Flatt\'e form\footnote{In the context of an effective field theory, the real $K$-matrix pole positions can be related to the bare masses of auxiliary fields associated with the resonances in the limit that these do not couple to scattering states. The coupling to asymptotic states dresses the auxiliary fields masses, giving them an imaginary component.}. An illustrative example of a parameterization of the type given by Eqn.~\ref{eq:K}, which proves to be successful in describing the spectra in Figure~\ref{fig:J2_spectra} is provided by
\begin{align}
K_{ij}(s) &= \frac{g^{(1)}_i g^{(1)}_j}{m_1^2 -s} + \frac{g^{(2)}_i g^{(2)}_j}{m_2^2 -s} + \gamma_{ij}, \nonumber \\[2ex]
&\quad\quad\boldsymbol{\gamma} = \begin{pmatrix} 0 & 0 & 0 \\ 0 & 0 & 0 \\ 0 & 0 & \gamma_{\eta\eta, \eta\eta} \end{pmatrix}, \label{eq:J2_ref_amp}
\end{align}
where in addition no real part is included in $\mathbf{I}$ (conventional phase-space). This amplitude features 9 real parameters: a mass and three channel couplings for each real pole plus a constant to allow for some gentle $\eta\eta$ energy dependence. This form allows the $\eta\eta$ channel to decouple if the fit selects small values for the relevant pole couplings ($g^{(1,2)}_{\eta\eta}$).

%%%%%%%%%%%%%%%%%%%%%%%%%%%%%%%%%%%%%%%%%%%%%%%%%%%%%%%%%%%%%%%%%%%%%
\begin{figure}
\hspace*{-4mm}\includegraphics[width = 1.05\columnwidth]{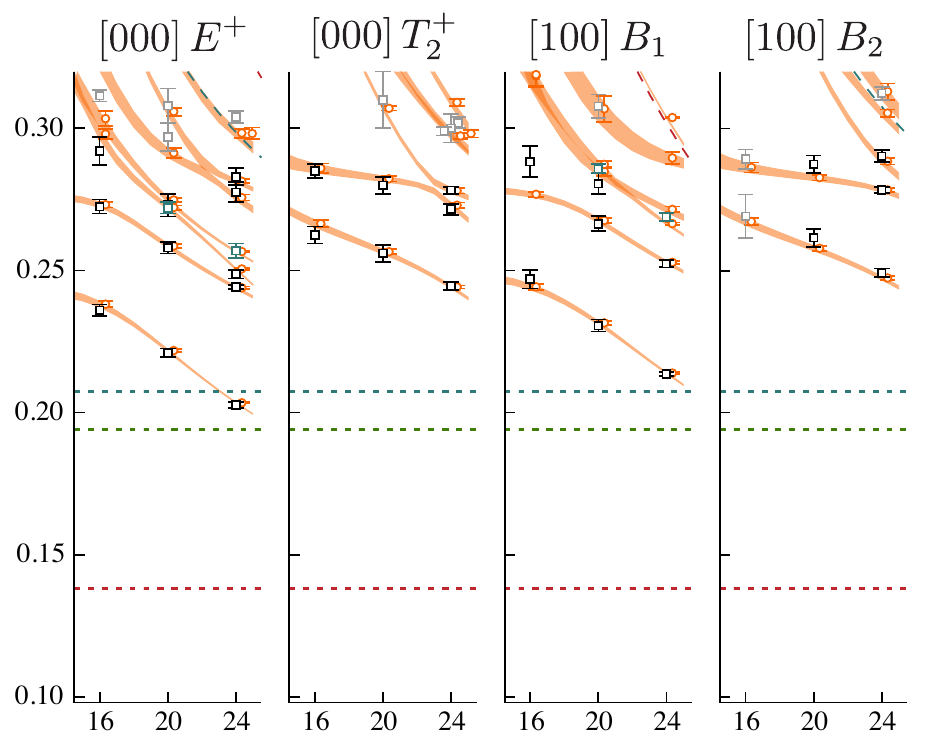}
\caption{Spectra in irreps with $\ell=2$ as lowest subduced partial wave, as described by the amplitude given in Eq.~\ref{eq:J2_ref_amp}. Black/blue points show the lattice QCD spectrum of Figure~\ref{fig:J2_spectra} and orange curves and points (displaced slightly for clarity) the result of the fitted amplitude. Long dashed lines indicate the position of non-interacting meson-meson levels corresponding to operators that were not included in the variational basis.}\label{fig:J2_model_spec}
\end{figure}
%%%%%%%%%%%%%%%%%%%%%%%%%%%%%%%%%%%%%%%%%%%%%%%%%%%%%%%%%%%%%%%%%%%%%

%%%%%%%%%%%%%%%%%%%%%%%%%%%%%%%%%%%%%%%%%%%%%%%%%%%%%%%%%%%%%%%%%%%%%
\begin{figure}
\hspace*{-7mm}
\includegraphics[width = 1.15\columnwidth]{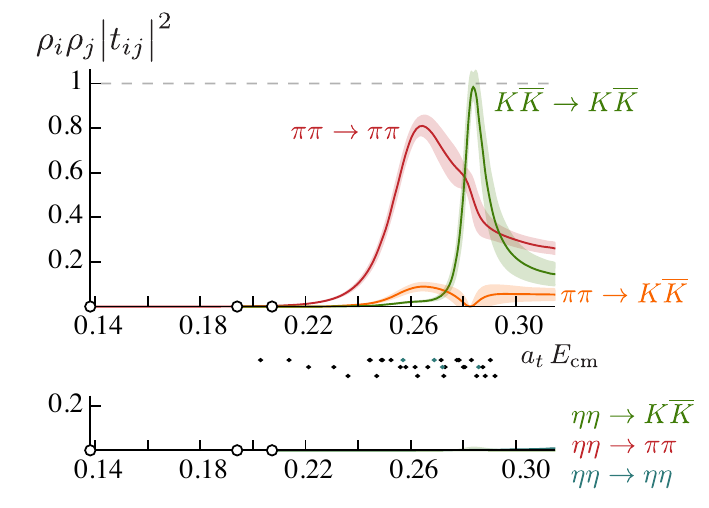}
\caption{The $D$-wave scattering amplitude of Eq.~\ref{eq:J2_ref_amp} plotted as in Figure~\ref{fig:k_inv_poly_llcccc_rho_t_sq}. }\label{fig:J2_ref_amp}
\end{figure}
%%%%%%%%%%%%%%%%%%%%%%%%%%%%%%%%%%%%%%%%%%%%%%%%%%%%%%%%%%%%%%%%%%%%%

%%%%%%%%%%%%%%%%%%%%%%%%%%%%%%%%%%%%%%%%%%%%%%%%%%%%%%%%%%%%%%%%%%%%%
\begin{figure*}
\includegraphics[width = 0.8\textwidth]{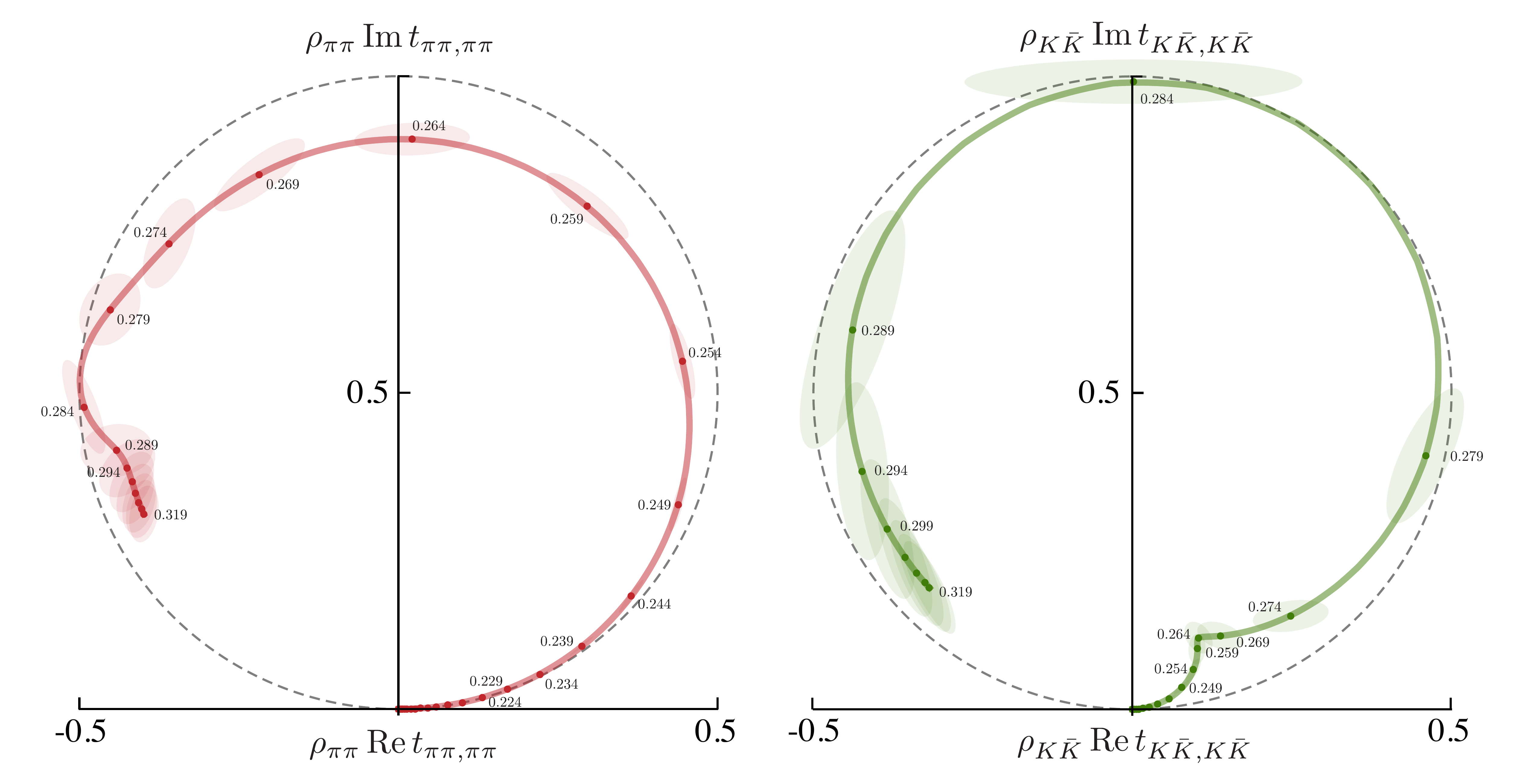}
\caption{Argand diagram representation of the $t_{\pi\pi, \pi\pi}$~(left) and $t_{K\overline{K}, K\overline{K}}$~(right) elements of the $D$-wave scattering amplitude given by Eq.~\ref{eq:J2_ref_amp}. Points are spaced evenly in energy with $a_t \Delta E_\mathrm{cm} = 0.005$.}\label{fig:J2_argand}
\end{figure*}
%%%%%%%%%%%%%%%%%%%%%%%%%%%%%%%%%%%%%%%%%%%%%%%%%%%%%%%%%%%%%%%%%%%%%

The description of the lattice QCD spectra provided by this amplitude, with a $\chi^2/N_\mathrm{dof} = 28.9/(34-9) = 1.15$, is shown in Figure~\ref{fig:J2_model_spec}. The description of the input spectrum is seen to be good, and furthermore the fit makes predictions for states which agree rather well with computed levels that were conservatively not included in the fit (ghosted in the figure).

Figure~\ref{fig:J2_ref_amp} shows the energy dependence of the resulting amplitudes, where we clearly observe significant enhancement in $\pi\pi$ slightly above $a_t E_\mathrm{cm} = 0.26$ and in $K\overline{K}$ slightly above $a_t E_\mathrm{cm} = 0.28$. Essentially no activity is seen in $\eta \eta$ as one would expect given the proximity of the relevant energy levels to the non-interacting $\eta\eta$ energies. The behavior observed in the $\pi\pi, K\overline{K}$ sector would appear to be that of two narrow resonances, the lighter decaying strongly to $\pi\pi$ and more weakly to $K\overline{K}$ and the heavier likely decaying mainly to $K\overline{K}$.

We note in passing the inactivity of the various components of the $\ell=2$ amplitude in the kinematic region where the $\ell=0$ were predominantly constrained in the previous section, namely  energies satisfying $a_t E_\mathrm{cm} < 0.24$. This explains why the contamination from the $\ell=2$ partial wave into the $A_1$ irreps played a negligible role in the analysis of the spectrum. 

It is interesting to examine the Argand diagrams for $\pi\pi \to \pi\pi$ and $K\overline{K} \to K\overline{K}$ for this $t$-matrix -- they are presented in Figure~\ref{fig:J2_argand}. The lower-energy bump has what appears to be a canonical resonance behavior in $\pi\pi \to \pi\pi$,  with the amplitude going counterclockwise through the vertical at $a_t E_\mathrm{cm} = 0.264$ and where the curve lying inside the unitarity circle is indicative of a loss of probability into the $K\overline{K}$ final state. The lower bump is visible in $K\overline{K} \to K\overline{K}$ as a strong kink at $a_t E_\mathrm{cm} = 0.264$, and above this energy the amplitude goes back onto the unitarity circle, suggesting an approximate decoupling from $\pi\pi$ at these energies. Another canonical resonance behavior near $a_t E_\mathrm{cm} = 0.284$ corresponding to the higher energy bump is present in $K\overline{K} \to K\overline{K}$.

Ultimately a rigorous approach to the resonance content and their relative decay strengths to open channels will come through consideration of the pole singularities of the \mbox{$t$-matrix} -- this will be discussed in Section~\ref{Sec:poles_D}.

%%%%%%%%%%%%%%%%%%%%%%%%%%%%%%%%%%%%%%%%%%%%%%%%%%%%%%%%%%%%%%%%%%%%%%%%%%%%%%%%%%%%%%%%%%%%%%%%%%%%%%%%%%%%%%%%%%%
\subsubsection{Varying the amplitude parameterization}

\begin{figure}
\hspace*{-8mm}
\includegraphics[width = 1.15\columnwidth]{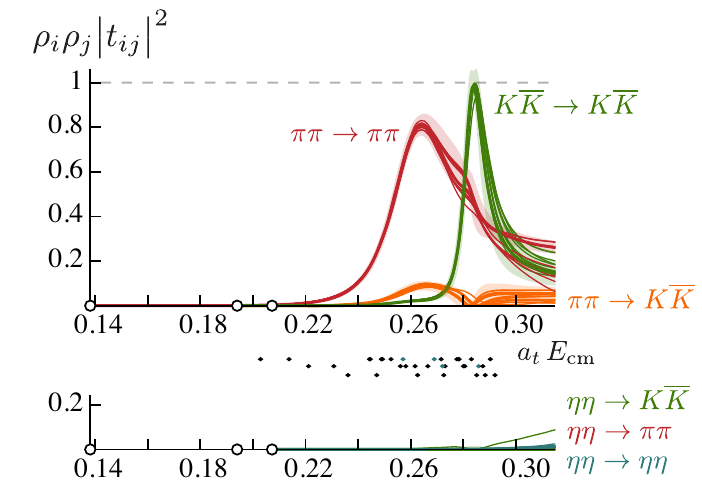}
\caption{Thick lines and bands show the amplitude previously plotted in Figure~\ref{fig:J2_ref_amp}. Thinner lines show the central value of amplitudes from 15 further parameterizations which successfully describe the lattice QCD spectra. }\label{fig:J2_amp_summary}
\end{figure}

The parameterization of Eq.~\ref{eq:J2_ref_amp} successfully describes the finite-volume spectra shown in Figure~\ref{fig:J2_spectra} in terms of two narrow bump structures. We find that only amplitudes featuring two such bumps are able to describe the spectrum, and in this section we present the result of considering a broader set of parameterizations, retaining two real poles in $K$, but adjusting some features: whether we allow the poles to couple to $\eta\eta$, the form of the quantity added to the poles, and the nature of $\mathbf{I}$ (Chew-Mandelstam versus naive phase-space). In total 16 amplitudes are presented
\footnote{full details are provided in Supplemental Material.}
, all of which have ${\chi^2/N_\mathrm{dof} < 1.2}$, and which do not feature overly large parameter correlations -- the resulting amplitudes are shown in Figure~\ref{fig:J2_amp_summary}. There is clearly very little variation under changes in the parameterization. We do observe a slight variation in the magnitude of the `shoulder' in $\pi\pi \to \pi\pi$ at $a_t E_\mathrm{cm} \sim 0.28$ and in the strength of $\pi\pi \to \pi\pi$ above the energy region where we have constrained the amplitudes ($a_t E_\mathrm{cm} > 0.30$). One particular amplitude parameterization finds a somewhat larger $\eta \eta \to K\overline{K}$ component (the lone visible green line in the lower pane), but this behavior begins outside the energy region where we have constrained the amplitudes.

In summary it appears that the $\pi\pi$, $K\overline{K}$ sector is rather well determined, in particular the position of the `bumps'. The $\eta\eta$ channel is less well constrained, but there are very strong hints that it is largely decoupled. In Section~\ref{Sec:poles_D} we will examine the pole singularity content of these amplitudes and propose a resonance interpretation.

%%%%%%%%%%%%%%%%%%%%%%%%%%%%%%%%%%%%%%%%%%%%%%%%%%%%%%%%%%%%%%%%%%%%%
%%%%%%%%%%%%%%%%%%%%%%%%%%%%%%%%%%%%%%%%%%%%%%%%%%%%%%%%%%%%%%%%%%%%%

% ref_amp_k_2poles_free_geta_xxxxxc_noCM		chisq = 1.15
% amp1_k_2poles_ccxcxc_noCM						chisq = 1.11
% amp2_k_2poles_ccxxxc							chisq = 1.07
% amp3_k_2poles_ccxxxc_noCM						chisq = 1.07
% amp4_k_2poles_cxxcxc_noCM						chisq = 1.07
% amp5_k_2poles_cxxxxc							chisq = 1.04
% amp6_k_2poles_cxxxxc_noCM						chisq = 1.03
% amp7_k_2poles_free_geta_cxxxxc_noCM			chisq = 1.12
% amp8_k_2poles_free_geta_xxcxxc_noCM			chisq = 1.19
% amp9_k_2poles_free_geta_xxxxxc				chisq = 1.18
% amp10_k_2poles_ssxxxs_noCM					chisq = 1.07
% amp11_k_2poles_sxxxxs_noCM					chisq = 1.04
% amp12_k_2poles_xxcxcc_noCM					chisq = 1.10
% amp13_k_2poles_xxxxxc							chisq = 1.09
% amp14_k_2poles_xxxxxc_noCM					chisq = 1.07
% amp15_k_2poles_xxxxxs_noCM					chisq = 1.06

%\clearpage
%%%%% POLES %%%%%
\pagebreak
\section{Resonance Poles \label{Sec:poles}}
% !TEX root = ../f0_paper.tex

%%%%% POLES intro %%%%    

The partial-wave amplitudes we have determined, like those extracted from experimental measurements, are evaluated at real values of the scattering energy, but an understanding of features of these amplitudes, such as peaks and cusps, comes from considering their singularity structure in the complex $s$-plane. As well as the branch points required by unitarity at the opening of each new channel, amplitudes can have pole singularities that we may identify with bound-states (if they lie on the real axis below threshold) and resonances (if they lie off the real axis). In the region of a pole singularity at $s = s_0$, the elements of the $t$-matrix behave like
\begin{equation}
t_{ij}(s) \sim \frac{c_i \, c_j}{s_0 -s }, \label{eq:pole}
\end{equation}
and the real and imaginary parts of the pole position are often identified with the mass and width of a resonance as $\sqrt{s_0} = m_R \pm \tfrac{i}{2} \Gamma_R$. The residue at the pole can be factorized into complex-valued \emph{couplings}, $c_i$, which indicate how strongly the resonance couples to scattering channels.

The presence of branch cuts associated with each new channel opening means that the complex $s$-plane is multi-sheeted. In single channel scattering there are two sheets which can be labelled by the sign of the imaginary part of the cm-frame scattering momentum, $k$. $\mathrm{Im}\, k > 0$ is called the \emph{physical sheet}, since it includes the region $s= E_\mathrm{cm}^2 + i \epsilon$ lying just above the real axis, where physical scattering occurs. Moving down from the real axis into the complex plane for any energy above threshold, we pass though the cut onto the \emph{unphysical sheet} where $\mathrm{Im} \, k < 0$. Resonance poles appear in complex conjugate pairs on the unphysical sheet, but usually only the pole in the lower half-plane is close to physical scattering. Complex poles cannot appear on the physical sheet as their presence would indicate a violation of causality.

In the case of multichannel scattering, the sheet structure becomes more complicated. For $n_\mathrm{chan}$ channels there is still a physical sheet on which all channel momenta have a positive imaginary part, but there are now $2^{n_\mathrm{chan}} - 1$ unphysical sheets. We will attempt to avoid confusion in nomenclature by labelling sheets by the sign of the imaginary part of the momentum in the three channels, ordered as $(\pi\pi, K\overline{K}, \eta\eta)$. For physical scattering between the $\pi\pi$ and $K\overline{K}$ thresholds, sheet $\mathsf(-,+,+)$, which we will also call sheet {\sf II}, is the nearest sheet to the scattering axis. Between the $K\overline{K}$ and $\eta \eta$ thresholds, sheet $\mathsf(-,-,+)$, which we will also call sheet {\sf III}, is closest, and above $\eta \eta$ threshold, it is sheet $\mathsf(-,-,-)$ which is closest. Some further discussion of sheet structure and pole positions in the context of two-channel scattering can be found in \cite{Dudek:2016cru}. A relevant observation is that a single resonance typically appears as a pole
\footnote{actually a complex-conjugate pair of poles, but we typically speak of this as one pole.}
 on several unphysical sheets (`mirror poles'), with shifts in position that are small if the resonance is dominantly coupled to only one scattering channel, but which can be large if coupled strongly to multiple channels\footnote{an illustrative presentation of this effect in the context of the Flatt\'e amplitude for two-channels will be given later in Section~\ref{Sec:scalars}.}.

Because the amplitudes we considered in Section~\ref{Sec:amps} are described by explicit forms, we can continue them to complex values of $s$, and search for pole singularities on all Riemann sheets. Uncertainties on the pole positions and couplings extracted from the residues can be estimated by propagating through the correlated uncertainties on the fitted amplitude parameters.

  \subsection{$S$-wave resonances and bound-states \label{Sec:poles_S}}
  % !TEX root = ../f0_paper.tex

%%%%% POLES in S-wave %%%%   

\subsubsection{$\sigma$ bound-state pole }%%%%%%%%%%%%%%%%%%%%%%%%%%%%%%%%%%%%%%%%%%%%%%%%%%%%%%%%%%%%%%%%%%
Our discussion in Section~\ref{Sec:elastic} concerning elastic $\pi\pi$ scattering leads us to expect that there is a bound-state pole singularity, lying below $\pi\pi$ threshold, that we may associate with a \emph{stable} $\sigma$ meson. Indeed in all successful descriptions of the lattice QCD spectrum, including all those presented in Section~\ref{Sec:coupled_S}, we find the amplitude features such a pole singularity, and we observe very little movement in its position with variation of parameterization form. A best estimate for its position, including in the uncertainty a conservative measure of the rather small degree of variation with parameterization form is $a_t\sqrt{s_0} = 0.1316(9)$. The main observed variation is a systematically lower mass (typically by roughly 0.0005) for parameterizations which use the ordinary phase-space for $\mathbf{I}$ rather than the Chew-Mandelstam form. The coupling to the $\pi\pi$ channel, $a_t\, c_{\pi\pi} = 0.092(4)$, shows very little variation under parameterization form, with the uncertainty dominated by the uncertainty on the energy levels. In multichannel fits, couplings of this state to the $K\overline{K}$ and $\eta\eta$ channels can be obtained from $t$-matrix residues at the pole, but the very large extrapolation below the relevant thresholds renders these numbers largely meaningless.

% eff_range order 2 : .1320(8), c = .094(3)
% eff_range order 1 : .1318(8), c = .082(3)

% ref_amp           : .1320(7), c = .094(3)
% amp1              : .1319(8), c = .095(3)
% amp2				: .1319(7), c = .095(3)
% amp3				: .1320(8), c = .096(3)
% amp4				: .1318(7), c = .096(3)
% amp5 				: .1311(8), c = .092(2) ** no CM **
% amp6				: .1314(8), c = .090(2) ** no CM **
% amp7				: .1311(9), c = .089(2) ** no CM **
% amp8				: .1313(8), c = .091(3) ** no CM **
% amp9				: .1317(8), c = .092(4)
% amp10				: .1317(8), c = .092(5)
% amp11				: .1317(8), c = .089(4)
% amp12				: .1310(7), c = .091(4) ** no CM **
% amp13				: .1312(10),c = .090(3) ** no CM **
% amp14				: .1318(8), c = .092(5)
% amp15				: .1317(7), c = .092(5)
% amp16				: .1317(7), c = .093(4)
% amp17				: .1319(7), c = .092(4)
% amp18				: .1312(8), c = .089(4) ** LO Adler **
% amp19				: .1311(11),c = .090(4) ** no CM **

This pole position and coupling differs slightly from that presented in~\cite{Briceno:2016mjc} -- as discussed earlier, it follows from fitting a set of energy levels that are not identical to those presented in that reference. Owing to the more cautious estimation of systematic errors in this paper, the result above should be considered to supersede the one in~\cite{Briceno:2016mjc}.

%%%%%%%%%%%%%%%%%%%%%%%%%%%%%%%%%%%%%%%%%%%%%%%%%%%%%%%%%%%%%%%%%%%%%%%%%%%%%%%%%%%%%%%%%%%%%%%%%%%%%%%
\subsubsection{$f_0$ resonance pole }%%%%%%%%%%%%%%%%%%%%%%%%%%%%%%%%%%%%%%%%%%%%%%%%%%%%%%%%%%%%%%%%%%
%%%%%%%%%%%%%%%%%%%%%%%%%%%%%%%%%%%%%%%%%%%%%%%%%%%%%%%%%%%%%%%%%%%%%%%%%%%%%%%%%%%%%%%%%%%%%%%%%%%%%%%

In Section~\ref{Sec:coupled_S} we presented suspicions that the sharp dip in $\pi\pi \to \pi\pi$ and rapid turn on of $K\overline{K}$ amplitudes at the $K\overline{K}$ threshold might be due to a nearby resonance -- this can be tested by analytically continuing the $t$-matrix into the complex plane, and searching for poles on all Riemann sheets. The consistent result of doing this for all successful parameterizations is the presence of a nearby pole on sheet $\mathsf{II(-,+,+)}$. This pole is partnered by a `mirror' pole on sheet $\mathsf{(-,+,-)}$, which we expect is of the same origin and appears at roughly the same energy due to the small coupling to $\eta \eta$. As an example, the amplitude described in Section~\ref{Sec:coupled_S}, given by Eq.~\ref{eq:ref_amp}, is found to have poles at
\begin{align}
\mathsf{II(-,+,+)} & \hspace{5mm}  a_t\sqrt{s_0} = 0.2131(69) - \tfrac{i}{2} 0.0391(128) \nonumber \\
\mathsf{  (-,+,-)} & \hspace{5mm}  a_t\sqrt{s_0} = 0.2236(44) - \tfrac{i}{2} 0.0318(140), \nonumber
\end{align}
and while there are also poles found on sheets $\mathsf{III(-,-,+)}$ and $\mathsf{(-,-,-)}$, they are rather far into the complex plane such that they will not have a significant impact on physical scattering\footnote{complete summaries of the pole content of this and other amplitudes can be found in Supplemental Material.}.

In~\cite{Dudek:2016cru} we discussed the possible manifestation of a sheet {\sf II} pole lying near the second threshold in the scattering amplitude for two strongly-coupled channels, which is almost exactly the scenario we are witnessing here. There we concluded that the scattering amplitude of the primary channel would have an asymmetric peak near the second threshold, which obviously differs from the `dip' behavior we see in the $\pi\pi$ amplitude in, for example, Fig.~\ref{fig:k_inv_poly_llcccc_rho_t_sq}. This dip is explained by the interference between the $\sigma$ and the $f_0$ poles, something that was not considered in the simple discussion presented in~\cite{Dudek:2016cru}.

\begin{figure}
\includegraphics[width = \columnwidth]{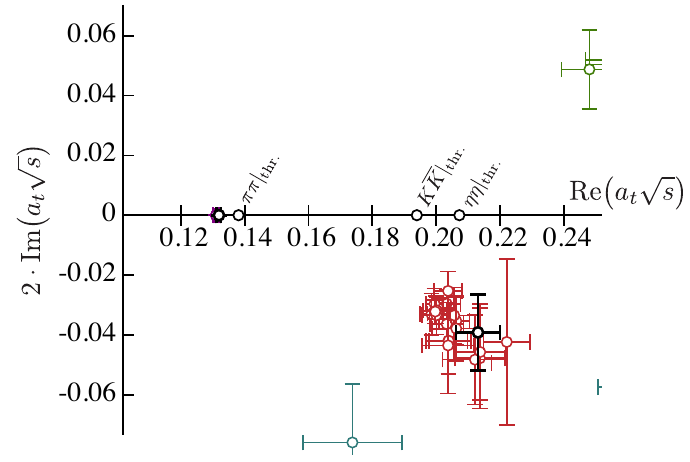}
\includegraphics[width = \columnwidth]{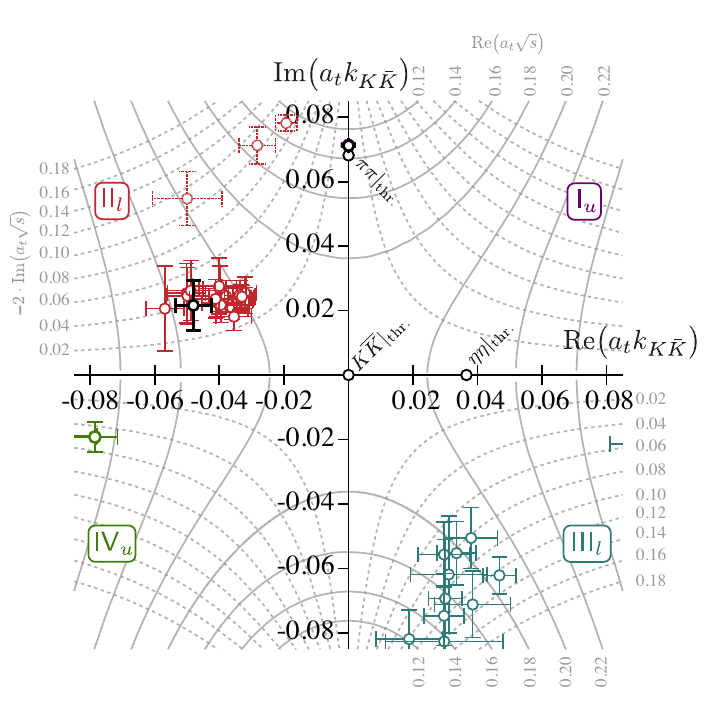}
\caption{Pole singularities for the 20 $S$-wave amplitudes discussed in Section~\ref{Sec:coupled_S}. Color indicates Riemann sheet of pole: sheet {\sf I}(purple), {\sf II}(red), {\sf III}(blue), and {\sf IV}(green). Thick black points indicate the particular amplitude defined by Eq.~\ref{eq:ref_amp}. Upper panel: complex $s$-plane. Lower panel: complex $k_{K\overline{K}}$-plane. Contours of constant complex energy plotted in lower panel to aid visualization of proximity of poles to physical scattering which occurs along the positive imaginary axis below $K\overline{K}$ threshold and along the positive real axis above the $K\overline{K}$ threshold. 
}\label{fig:J0_poles}
\end{figure}

%% variation %%
Figure~\ref{fig:J0_poles} shows the variation in pole location in the complex $s$-plane (upper panel) and complex \mbox{$k_{K\overline{K}}$-plane} (lower panel) for the 20 amplitudes presented in Section~\ref{Sec:coupled_S}, where we show only the poles on sheets {\sf I$(+,+,+)$}, {\sf II$(-,+,+)$}, {\sf III$(-,-,+)$} and {\sf IV$(+,-,+)$}. All these amplitudes feature a sheet {\sf II} pole at roughly the same location, while a few also have sheet {\sf III} or sheet {\sf IV} poles located further from physical scattering. A handful have a second sheet {\sf II} pole (visible in the lower panel as the dashed points) but again these are rather distant from physical scattering.

%%%%%%%%%%%%%%%%%%%%%%%%%%%%%%%%%%%%%%%%%%%%%%%%%%%%%%%%%%%%%%%%%%%%%
%%%%%%%%%%%%%%%%%%%%%%%%%%%%%%%%%%%%%%%%%%%%%%%%%%%%%%%%%%%%%%%%%%%%%

\begin{figure}
\includegraphics[width = \columnwidth]{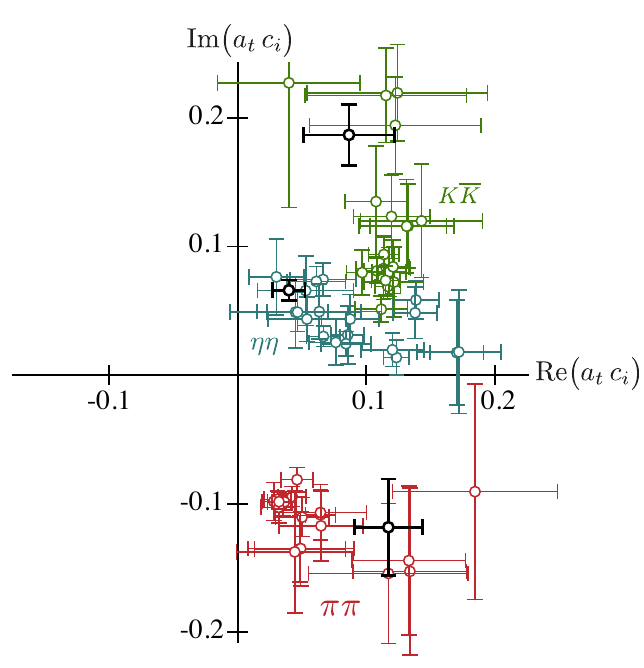}
\caption{Couplings for the 20 $S$-wave amplitudes discussed in Section~\ref{Sec:coupled_S} from factorized residues at the sheet {\sf II} pole. Thick black points indicate the particular amplitude defined by Eq.~\ref{eq:ref_amp}.} 
\label{fig:J0_couplings}
\end{figure}

Focussing on the consistent sheet {\sf II} pole, we may determine couplings to the $\pi\pi$, $K\overline{K}$ and $\eta \eta$ channels by factorizing the residues of $\mathbf{t}(s)$ at the complex pole position. The result of doing so is plotted in Figure~\ref{fig:J0_couplings}, where we observe $\pi\pi$ and $K\overline{K}$ couplings of comparable magnitudes, and a coupling to $\eta\eta$ that is somewhat smaller, albeit with some scatter under changes in parameterization.

%%%%%%%%%%%%%%%%%%%%%%%%%%%%%%%%%%%%%%%%%%%%%%%%%%%%%%%%%%%%%%%%%%%%%
%%%%%%%%%%%%%%%%%%%%%%%%%%%%%%%%%%%%%%%%%%%%%%%%%%%%%%%%%%%%%%%%%%%%%

%%%%%%%%%%%%%%%%%%%%%%%%%%%%%%%%%%%%%%%%%%%%%%%%%%%%%%%%%%%%%%%%%%%%%%%%%%%%%%%%%%%%%%%%%%%%%%%%%%%%%%
\subsubsection{The $K\overline{K}$ threshold region \label{sec:KKbar_thr}}  %%%%%%%%%%%%%%%%%%%%%%%%%%%%%%%%%%%%%%%%%%%%%%%%
%%%%%%%%%%%%%%%%%%%%%%%%%%%%%%%%%%%%%%%%%%%%%%%%%%%%%%%%%%%%%%%%%%%%%%%%%%%%%%%%%%%%%%%%%%%%%%%%%%%%%%

\begin{figure}
\hspace*{-6mm}\includegraphics[width = 1.05\columnwidth]{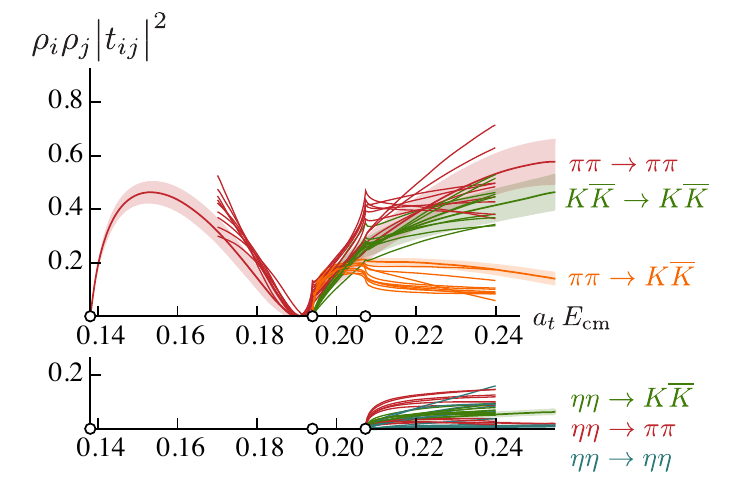}
\hspace*{-6mm}\includegraphics[width = 1.05\columnwidth]{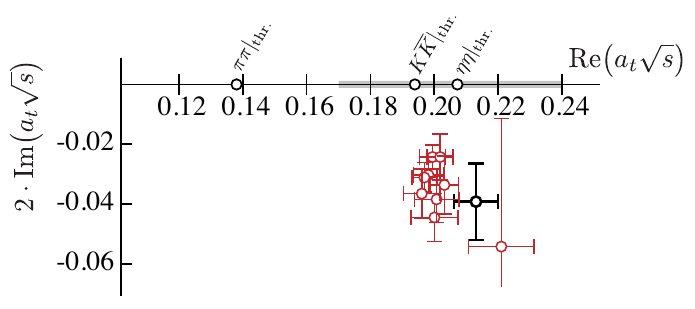}
\includegraphics[width = \columnwidth]{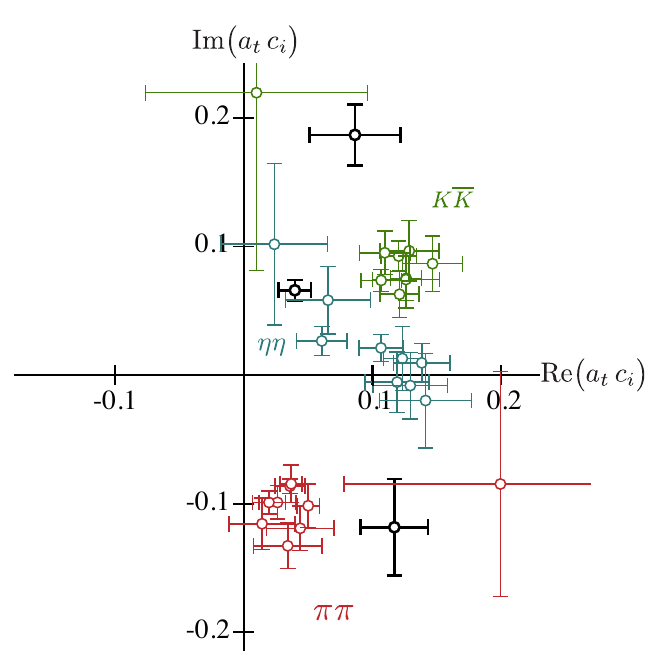}
\caption{Amplitudes describing the lattice QCD spectrum in the energy region $0.17 < a_t E_\mathrm{cm} < 0.24$. Top panel: Amplitudes. Middle panel: Sheet {\sf II} pole position for these amplitudes. Bottom panel: Couplings of sheet {\sf II} pole. Shown for comparison the amplitude described by Eq.~\ref{eq:ref_amp} (top panel -- curve with error bands, middle and bottom panels -- black points)}\label{fig:KKregion}
\end{figure}

The $f_0$ resonance pole described in the previous section dominates the amplitude in the energy region around the $K\overline{K}$ threshold, and in this region, the effect of the distant $\sigma$ bound-state is limited to providing a smoothly varying `background'. Given this, it is worthwhile to attempt to describe just that part of the spectrum lying above $a_t E_\mathrm{cm} = 0.17$ using amplitudes that need not lead to an explicit $\sigma$ bound-state pole. In Figure~\ref{fig:KKregion} we show 9 parameterizations which describe 41 levels in the energy region $0.17 < a_t E_\mathrm{cm} < 0.24$, all with $\chi^2/N_\mathrm{dof} < 1.05$. As in the previous section we note that the degree of coupling of the $\pi\pi, K\overline{K}$ sector to the $\eta\eta$ sector is somewhat imprecisely determined, but that otherwise there is very little variation in amplitude with change in parameterization. There is again always a sheet {\sf II} pole, but we note that it is systematically at a slightly lower mass than in the previous section. We observe there to be somewhat less scatter in the $\pi\pi$ and $K\overline{K}$ couplings, which show a small systematic shift in phase with respect to the previous section, but which have very similar magnitudes.

% amp1	k_inv_poly_cccccc		chisq = 1.03
% amp2	k_inv_poly_lccccc		chisq = 0.75
% amp3	k_inv_poly_ccclcc		chisq = 0.93
% amp4	k_inv_poly_cccccc_noCM	chisq = 0.81
% amp5	k_inv_poly_clcccc_noCM	chisq = 0.75
% amp6  k_inv_poly_ccclcc_noCM	chisq = 0.77
% amp7  k_poly_lccccc			chisq = 0.91
% amp8	k_poly_clcccc			chisq = 0.95
% amp9	k_poly_clcccc_noCM		chisq = 0.70

%%%%%%%%%%%%%%%%%%%%%%%%%%%%%%%%%%%%%%%%%%%%%%%%%%%%%%%%%%%%%%%%%%%%%
%%%%%%%%%%%%%%%%%%%%%%%%%%%%%%%%%%%%%%%%%%%%%%%%%%%%%%%%%%%%%%%%%%%%%

Note that some, but not all, of these amplitudes do feature a bound-state pole in roughly the position of the $\sigma$, but that this pole position is not precisely determined due to the energy levels below $a_t E_\mathrm{cm} = 0.17$ not being included in the fit. In those amplitudes which do not feature a bound-state pole, the effect of the $\sigma$ is being handled by smooth energy dependences that we might think of as `background'.

%%%%%%%%%%%%%%%%%%%%%%%%%%%%%%%%%%%%%%%%%%%%%%%%%%%%%%%%%%%%%%%%%%%%%%%%%%%%%%
\subsubsection{Controlling resonant pole content with Jost functions \label{sec:Jost}}  %%%%%%%
%%%%%%%%%%%%%%%%%%%%%%%%%%%%%%%%%%%%%%%%%%%%%%%%%%%%%%%%%%%%%%%%%%%%%%%%%%%%%%

As presented in the previous two sections, it appears that the lattice QCD spectra are best described by amplitudes which feature a sheet {\sf II} pole lying close to the $K\overline{K}$ threshold (as well as a bound-state $\sigma$ pole at much lower energy), and if they feature poles on sheet {\sf III} these are distant from physical scattering. This was all determined `after-the-fact', as the $K$-matrix forms used do not provide explicit control over the distribution of pole singularities. The position of the poles follows from a potentially complicated interplay of parameter values that is only determined once the fit is complete. On the other hand, Jost functions~\cite{Jost:1947,LeCouteur:1960,Newton:1961,Kato:1965} offer a parameterization of coupled-channel scattering in which the position of resonance poles in the multi-sheeted complex plane can be specified explicitly. We previously made use of such forms to describe two-channel $\pi\eta, K\overline{K}$ scattering in~\cite{Dudek:2016cru}, and indeed their implementation is much simplified for two-channel scattering, compared to three-channel scattering. Since we have found in previous sections that $\eta\eta$ appears to be weakly coupled, we choose to eliminate it here by excluding those levels identified previously as having large overlap onto $\eta \eta$ operators (those colored blue in Figure~\ref{fig:J0_spectra}). Furthermore, we avoid the complication of simultaneously describing the $\sigma$ and the $f_0$ by restricting our attention to those levels in the energy region, $0.17 < a_t E_\mathrm{cm} < 0.24$, around the $K\overline{K}$ threshold. We choose to make use of a conservative set of 30 levels.

\begin{figure}
\includegraphics[width = 0.92\columnwidth]{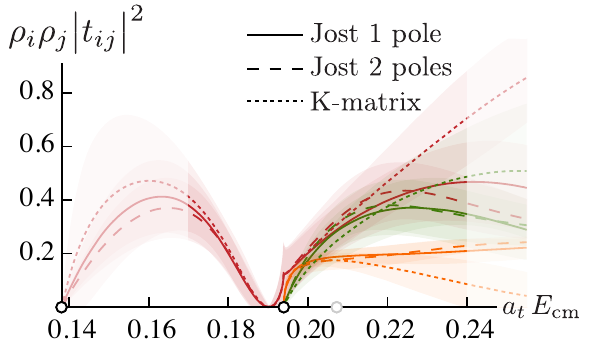}
\caption{Two-channel ($\pi\pi, K\overline{K}$) amplitudes using $K$-matrix and Jost-style parameterizations as is described in the text. The fits shown were obtained by requiring a description of energy levels in the region $0.17 < a_t E_\mathrm{cm} < 0.24$ excluding ``$\eta\eta$'' levels.}\label{fig:no_etaeta_fits}
\end{figure}

Before launching into a Jost function analysis, we first check that two-channel $K$-matrix analysis can successfully describe the spectrum with ``$\eta\eta$'' levels excluded. As an example, we find that a parameterization with symmetric $\mathbf{K}^{-1}$ having independent linear behavior ($a + b s$) in each element is able to describe the spectrum with ${\chi^2/N_\mathrm{dof} = 26.8/(30-6) = 1.12}$. This amplitude is shown in Figure~\ref{fig:no_etaeta_fits} where it is seen to closely resemble the $\pi\pi, K\overline{K}$ part of previous successful three-channel fits. This suggests that it is indeed reasonable to consider $\eta \eta$ to be decoupled.

The two-channel $S$-matrix is expressed in terms of the Jost determinant function $\mathfrak{J}$ by
\begin{equation*}
S_{11}=\frac{\mathfrak{J}(-k_1,k_2)}{\mathfrak{J}(k_1,k_2)},\; 
S_{22}=\frac{\mathfrak{J}(k_1,-k_2)}{\mathfrak{J}(k_1,k_2)},\; \det\mathbf{S}=\frac{\mathfrak{J}(-k_1,-k_2)}{\mathfrak{J}(k_1,k_2)}.
\end{equation*}
where $k_1$ and $k_2$ are the first and second channel cm-frame momenta ($k_1=k_{\pi\pi}$, $k_2=k_{K\overline{K}}$ in the current application). It is convenient to write this as a function of a single kinematic variable $\omega$, defined by
\begin{align}
\omega&=\frac{k_1+k_2}{\sqrt{k_1^2-k_2^2}}\,,\label{eq_omega}
\end{align}
as then the $S$-matrix elements can be expressed as
\begin{equation}
S_{11}=\frac{\mathfrak{D}(-\omega^{-1})}{\mathfrak{D}(\omega)},\;
S_{22}=\frac{\mathfrak{D}(\omega^{-1})}{\mathfrak{D}(\omega)},\; \det\mathbf{S}=\frac{\mathfrak{D}(-\omega)}{\mathfrak{D}(\omega)}.
\label{eq:JostSOmega}
\end{equation}
A convenient parameterization of $\mathfrak{D}(\omega)$, similar to that used in~\cite{Morgan:1993td}, features a product of zeroes, giving rise to $S$-matrix poles, multiplied by a smooth background function to describe the tail of the $\sigma$ bound-state, 
\begin{equation}
\mathfrak{D}(\omega) =  \exp \left( \sum_{b = 1}^{n_b} \gamma_b \, \omega^b \right) \, \frac{1}{\omega^2} \, \prod_{p=1}^{n_p} \left( 1 - \frac{\omega}{\omega_p}\right) \left( 1 + \frac{\omega}{\omega_p^*}\right). \label{eq:JostD}
\end{equation}
A constant term $\gamma_0$ does not appear since it cancels in the ratios in Eq.~\ref{eq:JostSOmega}. The real-analytic nature of the $S$-matrix implies that $\mathfrak{D}(\omega)=\mathfrak{D}^\star(-\omega^\star)$, which fixes $\mathrm{Re}(\gamma_{b_{\mathrm{odd}}})=0$ and $\mathrm{Im}(\gamma_{b_{\mathrm{even}}})=0$. This form does not contain any additional singularities in the energy region around $K\overline{K}$ threshold beyond the poles whose positions are specified by the values of the complex parameters $\omega_p$.

\begin{figure}
\includegraphics[width = \columnwidth]{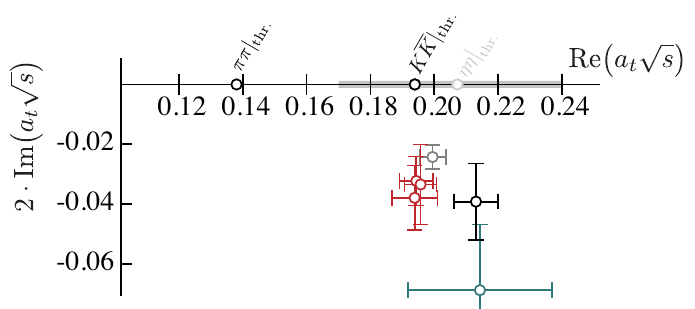}
\includegraphics[width = \columnwidth]{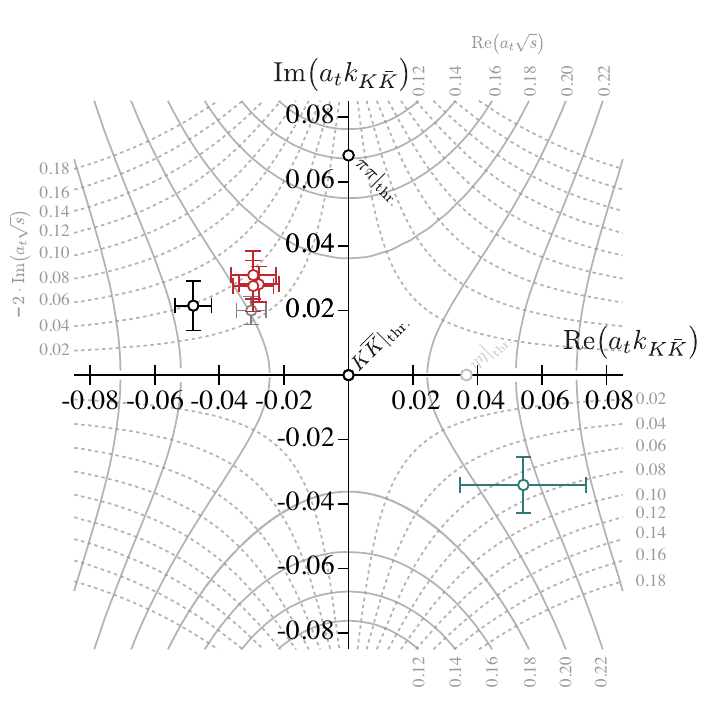}
\caption{Poles of two-channel ($\pi\pi, K\overline{K}$) amplitudes described in the text describing energy levels in the region $0.17 < a_t E_\mathrm{cm} < 0.24$ excluding ``$\eta\eta$'' levels. Black point shows the pole of the three-channel amplitude described by Eq.~\ref{eq:ref_amp}. Gray point shows the pole from a typical amplitude taken from Section~\ref{sec:KKbar_thr} where three-channel amplitudes described all levels in the region $0.17 < a_t E_\mathrm{cm} < 0.24$. Remaining colored points correspond to the amplitudes described in the text and plotted in Figure~\ref{fig:no_etaeta_fits}
}\label{fig:jost_poles}
\end{figure}

The results of previous sections lead us to believe that a parameterization with just a single pole may be capable of describing the spectra in the limited energy region around the $K\overline{K}$ threshold. An implementation of Eq.~\ref{eq:JostD} featuring one pole and with three terms in the background polynomial describes the spectrum with ${\chi^2/N_\mathrm{dof} = 28.8/(30-5) = 1.15}$. Figure~\ref{fig:no_etaeta_fits} shows the resulting amplitude which is quite similar to those we have previously seen. The position of the pole is allowed to float in the fit, and ends up in the expected position on sheet {\sf II}, as shown in Figure~\ref{fig:jost_poles}.

Attempting a description using Eq.~\ref{eq:JostD} with two poles and two terms in the exponentiated background polynomial, we find the spectrum is described with ${\chi^2/N_\mathrm{dof} = 29.4/(30-6) = 1.23}$. Having allowed both pole positions to float in the fit, we find one ends up in the expected position on sheet {\sf II}, while the other is poorly determined and appears to be distant on sheet {\sf III}. The amplitude is plotted in Figure~\ref{fig:no_etaeta_fits} and the poles are shown in Figure~\ref{fig:jost_poles}.

%couplings for the sheet III pole much smaller than for the sheet II pole

% jost_2pole_order2
% II   0.1944(53) -  (i/2)*0.0322(82)        	|c_pipi| = 0.095(17)	|c_KK| = 0.146(36)
% III  0.2143(226) - (i/2)*0.0687(219)			|c_pipi| = 0.079(24)    |c_KK| = 0.053(16)

As in our previous three-channel analysis, we observe that restricting the energy region under consideration to be around the $K\overline{K}$ threshold, discarding energy levels at lower energy, tends to cause the sheet {\sf II} pole to move to a slightly lower mass.

%%%%%%%%%%%%%%%%%%%%%%%%%%%%%%%%%%%%%%%%%%%%%%%%%%%%%%%%%%%%%%%%%%%%%
%%%%%%%%%%%%%%%%%%%%%%%%%%%%%%%%%%%%%%%%%%%%%%%%%%%%%%%%%%%%%%%%%%%%%

%%%%%%%%%%%%%%%%%%%%%%%%%%%%%%%%%%%%%%%%%%%%%%%%%%%%%%%%%%%%%%%%%%%%%
%%%%%%%%%%%%%%%%%%%%%%%%%%%%%%%%%%%%%%%%%%%%%%%%%%%%%%%%%%%%%%%%%%%%%

In Appendix~\ref{Sec:Jost_scan} we consider what happens if we force the amplitude to feature a sheet {\sf III} pole as well as the sheet {\sf II} pole demanded by the spectrum. As we move the sheet {\sf III} pole close to physical scattering, the amplitude comes to have a rapid energy dependence which is not supported by the lattice QCD spectra, leading to an unacceptably large $\chi^2$. For a more distant sheet {\sf III} pole, the background part of the amplitude parameterization is able to largely compensate, and the $\chi^2$ remains acceptable.

\clearpage
%%%%%%%%%%%%%%%%%%%%%%%%%%%%%%%%%%%%%%%%%%%%%%%%%%%%%%%%%%%%%%%%%%%%%%%%%%%%%%%%%%%%%%%%%
\subsubsection{$S$-wave resonance pole summary}    %%%%%%%%%%%%%%%%%%%%%%%%%%%%%%%%%%%%%%%%%%%%%%%
%%%%%%%%%%%%%%%%%%%%%%%%%%%%%%%%%%%%%%%%%%%%%%%%%%%%%%%%%%%%%%%%%%%%%%%%%%%%%%%%%%%%%%%%%

All successful descriptions of the lattice QCD spectra either over a large energy region, or restricted to the region around the $K\overline{K}$ threshold, feature a pole on sheet $\mathsf{II(-,+,+)}$ at roughly the same position. This pole is found to have large couplings to both $\pi\pi$ and $K\overline{K}$. A mirror pole on sheet $\mathsf{(-,+,-)}$ is also present. The lattice QCD spectra can tolerate in addition a fairly distant pole on sheet $\mathsf{III(-,-,+)}$, but if present it does not appear to be a dominant feature in the amplitude. 

Our best estimate for the properties of the sheet {\sf II} pole are
\begin{align}
a_t \sqrt{s_0} &= 0.2060(80) - \tfrac{i}{2} 0.032(12) \nonumber \\[0.8ex]
a_t |c_{\pi \pi}| &= 0.125(25) \nonumber\\
a_t |c_{K \overline{K}}| &= 0.150(20) \nonumber\\
a_t |c_{\eta\eta}| &= 0.090(35), \label{eq:f0pole}
\end{align}
where the quoted uncertainties take into account the variation over parameterization presented in this section.

  \subsection{$D$-wave resonances\label{Sec:poles_D}}
  % !TEX root = ../f0_paper.tex

%%%%% POLES in D-wave %%%%    

The pole content of the $D$-wave amplitude can be guessed quite easily from the energy dependence displayed in Figure~\ref{fig:J2_amp_summary}. In order to have such sharp peaks, there must be pole singularities on nearby sheets, and above all three thresholds ($\pi\pi$, $K\overline{K}$ and $\eta\eta$), the nearest unphysical sheet is $\mathsf{(-,-,-)}$. Hence we would expect there to be two poles on this sheet, with the one having lower mass being somewhat further from the real axis, corresponding to the larger width of the lower peak.

\begin{figure}
\includegraphics[width = \columnwidth]{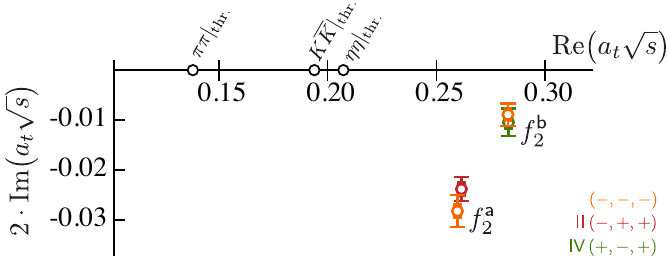}
\caption{Pole singularities for the $D$-wave amplitude of Eq.~\ref{eq:J2_ref_amp}. As well as the sheets shown, there are also poles on sheet $\mathsf{III(-,-,+)}$ lying at almost exactly the same position as the sheet $\mathsf{(-,-,-})$ poles.}\label{fig:J2_poles_ref_amp}
\end{figure}

Indeed, when we examine the pole content of the amplitude described in Section~\ref{Sec:coupled_D}, presented in Eq.~\ref{eq:J2_ref_amp}, we find two poles on sheet $\mathsf{(-,-,-)}$. As shown in Figure~\ref{fig:J2_poles_ref_amp}, these poles (shown in orange) have `mirrors' on other unphysical sheets. The lower mass pole, which dominantly couples to $\pi\pi$, and which we will label ``$\,f_2^\mathsf{a}\,$'', has mirrors on sheet {\sf II} and sheet {\sf III} (not visible in the plot as it lies almost exactly underneath the $\mathsf{(-,-,-)}$ pole). The higher mass pole, which couples dominantly to $K\overline{K}$, and which we will label ``$\, f_2^\mathsf{b}\, $'', has mirrors on sheet {\sf IV} and sheet {\sf III} (not visible in the plot as it lies almost exactly underneath the $\mathsf{(-,-,-)}$ pole). The mirror poles lie at almost the same position owing to the relatively small couplings to sub-dominant channels.

\begin{figure}
\includegraphics[width = \columnwidth]{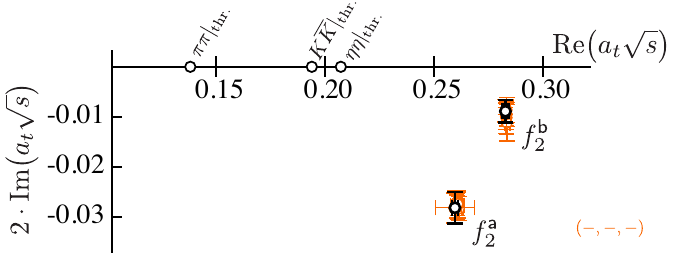}
\caption{Variation of $D$-wave sheet $\mathsf{(-,-,-)}$ poles with amplitude parameterization as described in Section~\ref{Sec:coupled_D}.}\label{fig:J2_poles_summary}
\end{figure}

\begin{figure}
\includegraphics[width = 0.85\columnwidth]{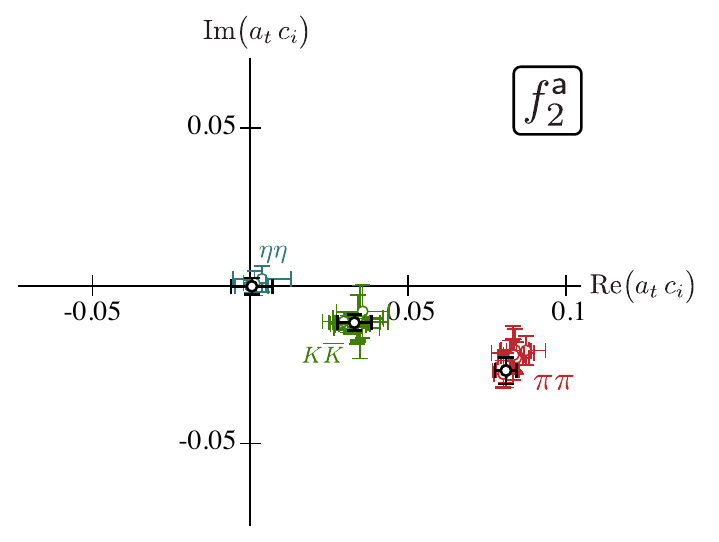}
\includegraphics[width = 0.85\columnwidth]{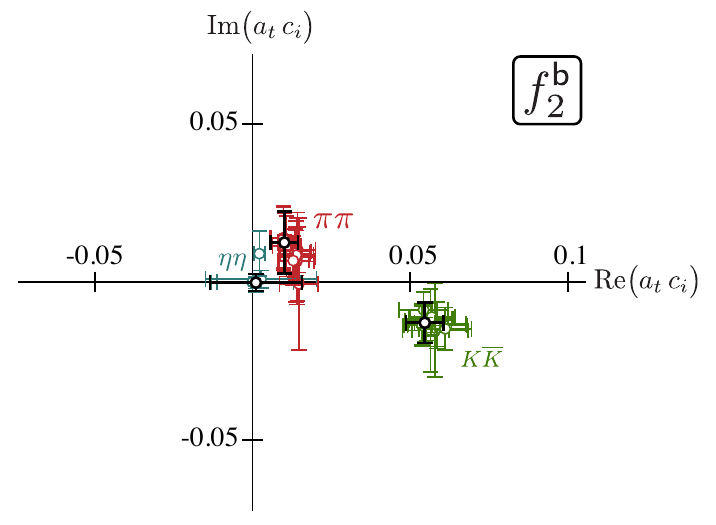}
\caption{Couplings from factorized pole residues for two sheet $\mathsf{(-,-,-)}$ poles, $f_2^\mathsf{a}$, $f_2^\mathsf{b}$. Black points show couplings for the amplitude of Eq.~\ref{eq:J2_ref_amp}, others show the variation with change in parameterization described in Section~\ref{Sec:coupled_D}.}\label{fig:J2_couplings}
\end{figure}

Focussing on the $\mathsf{(-,-,-)}$ poles, since these are the closest to physical scattering, we show the variation with parameterization presented in Section~\ref{Sec:coupled_D} in Figure~\ref{fig:J2_poles_summary}, which we observe to be extremely small. Figure~\ref{fig:J2_couplings} shows the couplings extracted from the residues of poles on the $\mathsf{(-,-,-)}$ sheet. Again, the variation with parameterization change is extremely small, and in all cases the $\eta \eta$ coupling is compatible with zero. As one would expect from Figure~\ref{fig:J2_amp_summary}, the $\pi\pi$ coupling dominates for $f_2^\mathsf{a}$, and the $K\overline{K}$ coupling dominates for $f_2^\mathsf{b}$.

%%%%%%%%%%%%%%%%%%%%%%%%%%%%%%%%%%%%%%%%%%%%%%%%%%%%%%%%%%%%%%%%%%%%%
%%%%%%%%%%%%%%%%%%%%%%%%%%%%%%%%%%%%%%%%%%%%%%%%%%%%%%%%%%%%%%%%%%%%%

%%%%%%%%%%%%%%%%%%%%%%%%%%%%%%%%%%%%%%%%%%%%%%%%%%%%%%%%%%%%%%%%%%%%%
%%%%%%%%%%%%%%%%%%%%%%%%%%%%%%%%%%%%%%%%%%%%%%%%%%%%%%%%%%%%%%%%%%%%%

%%%%%%%%%%%%%%%%%%%%%%%%%%%%%%%%%%%%%%%%%%%%%%%%%%%%%%%%%%%%%%%%%%%%%

Our best estimates for the properties of the two poles on sheet $\mathsf{(-,-,-)}$ are
\begin{align*}
&f_2^\mathsf{a}:\;\;\; a_t \sqrt{s_0} = 0.2596(26) - \tfrac{i}{2} 0.0282(32) \nonumber \\
&a_t  \big|c_{\pi\pi} \big| = 0.086(5),\; a_t  \big|c_{K\overline{K}} \big| = 0.036(7),
\end{align*}
\begin{align*}
&f_2^\mathsf{b}:\;\;\; a_t \sqrt{s_0} = 0.2829(17) - \tfrac{i}{2} 0.0095(25) \nonumber \\
&a_t  \big|c_{\pi\pi} \big| = 0.016(5),\; a_t  \big|c_{K\overline{K}} \big| = 0.058(9),
\end{align*}
where the uncertainties include the small variation with parameterization form described above.

%\clearpage
%%%%% INTERPRETATION %%%%%
\section{Interpretation \label{Sec:interpret}}
% !TEX root = ../f0_paper.tex

%%%%% INTERPRET %%%%    

%%%%%%%%%%%%%%%%%%%%%%%%%%%%%%%%%%%%%%%%%%%%%%%%%%%%%%%%%%%%%%%%%%%%%
\begin{figure*}
\hspace*{-8mm}\includegraphics[width = 1.1\textwidth]{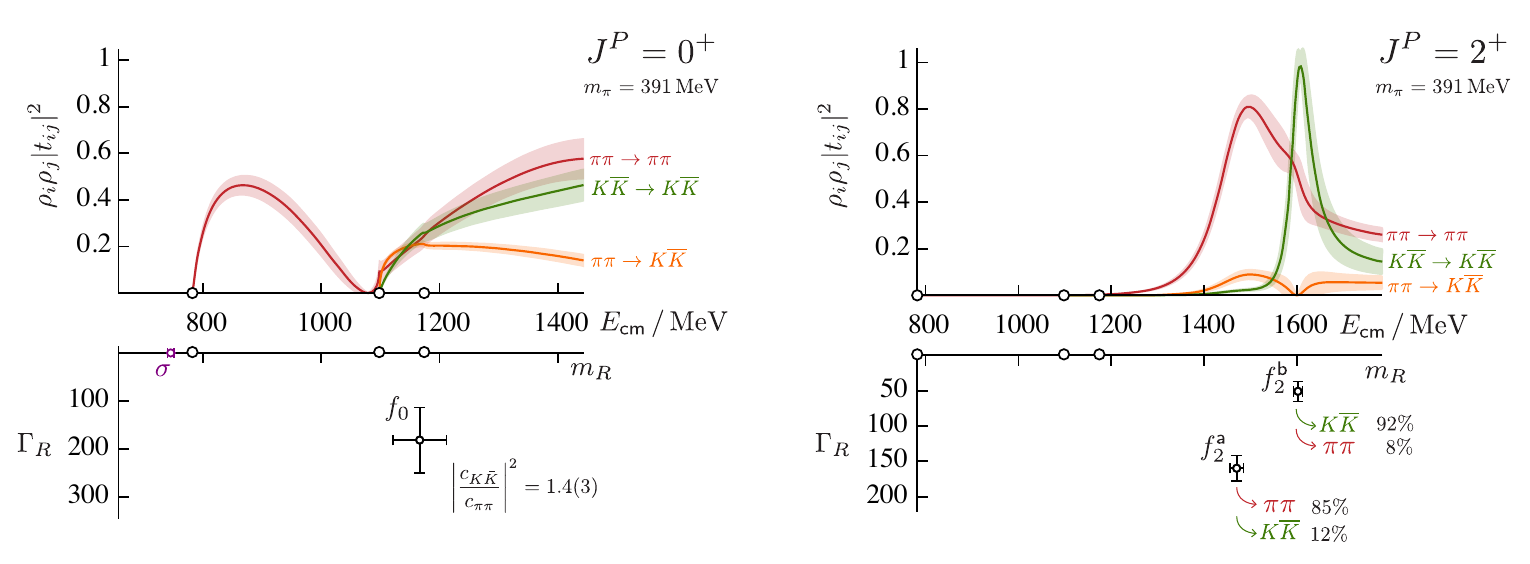}
\caption{Coupled $\pi\pi,K\overline{K}$ amplitudes obtained using Eq.~\ref{eq:ref_amp} for the scalar (left) and Eq.~\ref{eq:J2_ref_amp} for the tensor (right) sectors. In both cases, the $\eta\eta$ channels are approximately decoupled. Also shown are the pole singularities, with uncertainties which include the variation under amplitude parameterization presented in Sections~\ref{Sec:coupled_S},~\ref{Sec:coupled_D}. The ratio of couplings to $K\overline{K}, \pi\pi$ for the $f_0$ is presented, as are estimates of the branching fractions for the $f_2$ resonances, as described in the text.  }\label{fig:scale_set}
\end{figure*}
%%%%%%%%%%%%%%%%%%%%%%%%%%%%%%%%%%%%%%%%%%%%%%%%%%%%%%%%%%%%%%%%%%%%%

As in previous publications~\cite{Dudek:2012gj, Dudek:2012xn, Dudek:2014qha, Wilson:2014cna, Dudek:2016cru, Briceno:2016kkp, Briceno:2016mjc, Moir:2016srx}, we opt to set the lattice scale using the $\Omega$ baryon mass computed on these lattices, setting it equal to the experimentally measured mass~\cite{Olive:2016xmw}, which yields $a_t^{-1} = 5.662 \,\mathrm{GeV}$. With this scale-setting the stable $\pi,K, \eta$ masses are approximately $391, 549, 587$~MeV respectively. 

In previous sections we have described a study of coupled $\pi\pi, K\overline{K}, \eta \eta$ scattering with isospin=0 ($G$-parity positive). With $J^P=0^+$ we found two singularities are required: a bound-state $\sigma$ pole appearing on the physical sheet at $\sqrt{s_0} = 745(5)$~MeV with a coupling to the $\pi\pi$ channel of $|c_{\pi\pi}| = 521(23)$~MeV, and a resonance $f_0$ pole on sheet {\sf II} at $\sqrt{s_0} = 1166(45) - \tfrac{i}{2}\, 181(68)$~MeV, with couplings $|c_{\pi\pi}| = 710(140)$~MeV, and $|c_{K\overline{K}}| = 850(110)$~MeV.

The $\sigma$ dominates low-energy $\pi\pi$ elastic scattering, giving rise to an $S$-wave scattering length of $m_\pi \, a = -2.8(3)$. As reported on in~\cite{Briceno:2016mjc}, as the light quark mass is reduced, this meson evolves from being a stable bound-state into a broad resonance.

%%% summary over amplitudes
% m_pi * a

% elastic a,r 		-2.53(22)
% elastic a,r,P2	-2.98(27)

% ref_amp			-3.00(20)
% amp1				-2.98(20)
% amp2				-2.98(21)
% amp3				-3.06(21)
% amp4				-2.98(20)
% amp5				-2.68(20)
% amp6				-2.68(21)
% amp7				-2.61(21)
% amp8				-2.70(23)
% amp9 				-2.86(24)
% amp10				-2.87(27)
% amp11				-2.75(24)
% amp12				-2.64(23)
% amp13				-2.64(28)
% amp14				-2.87(27)
% amp15				-2.85(23)
% amp16				-2.88(23)
% amp17 			-2.90(25)
% amp18				-2.64(21)
% amp19				-2.62(30)

The $f_0$ resonance, lying in the $K\overline{K}$ threshold region, has not previously been observed in a first-principles QCD calculation. It appears to share many of the properties of the experimental $f_0(980)$ resonance, and we suggest that what we are observing may well represent the evolution of this meson as the light quark mass increases. 
The absence of a nearby sheet {\sf III} mirror to the $f_0$ \mbox{sheet {\sf II} pole} appears to support a longstanding suggestion that the $f_0$ resonance may be dominated by $K\overline{K}$-molecule configurations (see for example Refs.~\cite{Morgan:1992ge, Morgan:1993td, Baru:2004xg}). The logic is that a $K\overline{K}$ molecular state bound by long-range inter-meson forces would be a stable bound-state lying just below the $K\overline{K}$ threshold were it not for the kinematically open $\pi\pi$ channel into which it decays. The presence of such a decay moves the pole off the real energy axis into sheet {\sf II}. This is to be compared to a compact state bound by confining interquark forces (whether $q\bar{q}$ or tetraquark or an even higher quark-gluon Fock state), which is expected to manifest itself as `mirror' poles on both of sheets {\sf II} and {\sf III}. 

Presently, the absence of a formalism describing scattering of more than two hadrons prevents us from considering higher-mass scalar meson resonances. Experimentally, these are seen to have dominant decays to four-meson final states. Fortunately, the formalism has been under rapid development~\cite{Briceno:2017tce, Hammer:2017kms, Hansen:2014eka, Hansen:2015zga, Briceno:2012rv, Polejaeva:2012ut}, and it is hoped that a final result for at least three-body decays will be available shortly.  

%%% J=2 %%

With $J^P = 2^+$ we isolated the presence of two narrow resonances lying well above the $\pi\pi$, $K\overline{K}$ and $\eta \eta$ thresholds, but with negligible couplings to the $\eta \eta$ channel. Because they lie significantly above thresholds, it makes sense to speak of `branching fractions' for their decay. We compute these using the approach outlined by the PDG \cite{Olive:2016xmw}, where the real and imaginary parts of the pole position, and the couplings extracted from the factorized residue at the pole are used in
\begin{equation*}
\mathrm{Br}(R \to i) = \frac{1}{\Gamma_R}\cdot  \frac{|c_i|^2}{m_R} \rho_i(m_R^2).
\end{equation*}
We note in passing that this approach does not guarantee that the sum of branching fractions be 100\%. Our two determined resonances have the following properties: 
\begin{align*}
&f_2^\mathsf{a}:\;\;\; \sqrt{s_0} = 1470(15) - \tfrac{i}{2} \, 160(18)\;\mathrm{MeV} \nonumber \\
&\mathrm{Br}(f_2^\mathsf{a} \to \pi\pi) \sim 85\%, \;\;\; \mathrm{Br}(f_2^\mathsf{a} \to K\overline{K}) \sim 12\% \,,
\end{align*}
\begin{align*}
&f_2^\mathsf{b} :\;\;\;  \sqrt{s_0} = 1602(10) - \tfrac{i}{2} \, 54(14)\; \mathrm{MeV} \nonumber \\ 
&\mathrm{Br}(f_2^\mathsf{b} \to \pi\pi) \sim 8\%, \;\;\; \mathrm{Br}(f_2^\mathsf{b} \to K\overline{K}) \sim 92\% \,.
\end{align*}

These resonances, computed with heavier than physical light quarks, may be compared to the experimental resonances, $f_2(1270)$ and $f_2'(1525)$ \cite{Olive:2016xmw}. The experimental $f_2(1270)$ has a total width $\sim 190$~MeV and decays 84\% of the time to $\pi\pi$ and only 5\% to $K\overline{K}$. The $f_2'(1525)$ is narrower, $\Gamma \sim 80$~MeV, and decays to $K\overline{K}$ with a 90\% branch, and to $\pi\pi$ less than 1\% of the time.

These states are often considered to be exemplars of the phenomenological `OZI' rule of meson decays, which posits that decays proceeding through annihilation of existing quark-antiquark content are suppressed with respect to decays in which extra quarks are generated. In this case this would suggest that $f_2(1270) \sim \tfrac{1}{\sqrt{2}}\big( u\bar{u} + d \bar{d} \big)$ with the decays to $\pi\pi$ and $K\overline{K}$ being `OZI-allowed' through creation of extra light or strange quark-antiquark pairs respectively
\footnote{with $K\overline{K}$ suppressed with respect to $\pi\pi$ by the reduced \mbox{phase-space} and additionally possibly a dynamical penalty for \mbox{pair-producing} heavier $s\bar{s}$ quarks.}
, while $f_2'(1525) \sim s\bar{s}$ will decay only to $K\overline{K}$ through creation of extra light quark-antiquark pairs, with $\pi\pi$ requiring the initial $s\bar{s}$ to annihilate -- an `OZI-suppressed' decay. This logic can be turned around and used to infer a resonance's quark content on the basis of its preferred hadronic decays.

It is interesting to observe that the results of our calculation (at $m_\pi \sim 391$~MeV) appear to support this picture with the $f_2^\mathsf{a}$ having a decay pattern like the $f_2(1270)$, while the $f_2^\mathsf{b}$ closely resembles the $f_2'(1525)$. A somewhat non-rigorous suggestion for the quark-antiquark content of these two states can be obtained by examining the overlaps presented in Figure~\ref{fig:spec_000_J2_histo}, in particular the $16^3$ spectrum in the $T_2^+$ irrep, where two finite-volume states are observed, far from any non-interacting energy levels, but close to the resonance masses for $f_2^\mathsf{a}$ and $f_2^\mathsf{b}$. The lighter of the two is seen to have large overlap onto $\bar{q}\mathbf{\Gamma}q$ operator constructions built from light-quarks, while the heavier state dominantly overlaps with those made from strange-quarks. 

We note in passing that the presence of two narrow $J^P=2^+$ resonances with this quark content was anticipated in the simpler calculation of the isoscalar meson spectrum presented in \cite{Dudek:2013yja}, where no meson-meson-like operators where included. Some discussion of the justification for expecting the results presented in~\cite{Dudek:2013yja} to be a reasonable guide to the narrow resonance content of QCD is presented in~\cite{Briceno:2017max}.

Experimentally, the $f_2(1270)$ is observed to have negligible coupling to $\eta \eta$, and only a $\sim 10\%$ branch to $\pi\pi\pi\pi$, which is not kinematically open at $m_\pi \sim 391$~MeV. The $f_2'(1525)$ has a $\sim 10\%$ branch to $\eta \eta$ -- this small but significant coupling does not appear to be present at the heavier light quark mass considered in this paper, despite the phase-space for the decay being quite similar to the experimental case.

A plausible method to more rigorously study the internal quark-gluon structure of resonance states of the type we have considered here is to compute their form-factors, by coupling external currents into meson-meson scattering amplitudes. The possibility of studying form-factors of unstable states from lattice QCD was recently reviewed in~\cite{Briceno:2017max}, building on ideas first presented by Lellouch and L\"uscher~\cite{Lellouch:2000pv} for the process $K\to\pi\pi$. The `elastic' form-factors of resonant states could be extracted from amplitudes computed at real energies, by extrapolating to the complex pole positions~\cite{Briceno:2015tza, Bernard:2012bi}. Exploiting the flexibility of lattice QCD calculation, one could study flavor-- and spin--dependence of these form factors, for example by introducing  $\bar{s}\Gamma s$ and $\bar{u} \Gamma u + \bar{d} \Gamma d $ currents separately, for various choices of $\Gamma$. The \emph{distillation} technology for such calculations has already been developed~\cite{Shultz:2015pfa}, and the first nontrivial test has been carried out in the resonant $\pi\gamma\to\pi\pi$ amplitude~\cite{Briceno:2015dca, Briceno:2016kkp} using the formalism first presented in~\cite{Briceno:2014uqa} for transition processes.

Fig.~\ref{fig:scale_set} presents a summary of the main results of the calculation reported on in this paper. The scalar and tensor $\pi\pi, K\overline{K}$ amplitudes are plotted, along with the corresponding pole structure, and determinations of the relative strengths of coupling of the resonances to their decay channels.

% for completeness the pipi scattering length for J=2 is
% ref_amp 			m_pi^5 * a = 0.0246(23) 

%%%%% SCALAR MESONS %%%%%
\section{The lightest scalar mesons of QCD at $m_\pi \sim 400$ MeV \label{Sec:scalars}}
% !TEX root = ../f0_paper.tex

%%%%% SCALARS %%%%    

With the results for isoscalar mesons presented in this paper, taken together with the results in Refs.~\cite{Dudek:2016cru, Wilson:2014cna, Dudek:2014qha} for isovector and strange mesons, we have what could constitute a complete nonet of scalar ($\sigma, f_0, a_0, \kappa$) and tensor ($f_2, f_2', a_2, K_2^\star$) mesons. It is appropriate at this stage to consider to what extent they have common properties that justifies associating them in this manner.

In the scalar sector, the two states which appear to be mostly closely connected are the $f_0$ and the $a_0$. While their appearance in the relevant elastic amplitude (${\pi\pi \to \pi\pi}$ or ${\pi \eta \to \pi \eta}$) is superficially rather different, being a sharp cusp-like peak for the $a_0$ and a narrow dip for the $f_0$, both effects appear close to the $K\overline{K}$ threshold, and the corresponding resonance poles are found at positions whose real parts are in close agreement,
\begin{align*}
&m_R(f_0) = 1166(45)\, \mathrm{MeV}, &\Gamma_R(f_0) &= 181(68)\, \mathrm{MeV}, \nonumber \\
&m_R(a_0) = 1177(27)\, \mathrm{MeV},  &\Gamma_R(a_0) &= 49(33)\, \mathrm{MeV}.
\end{align*}
In addition, the couplings of the resonances to their decay channels agree within statistical uncertainties,
\begin{align}
\big|c(a_0 \to K\overline{K})\big| &\approx \big|c(f_0 \to K\overline{K}) \big| &&\sim 850\,\mathrm{MeV} \nonumber \\
\big|c(a_0 \to \pi \eta)\big| &\approx \big|c(f_0 \to \pi\pi) \big| &&\sim 700 \,\mathrm{MeV}. \label{eq:f0a0_couplings}
\end{align}
Because isospin is an exact symmetry in our calculations, and the effects of electromagnetism are not included, there can be no mixing between the $a_0$ and the $f_0$.

Where the pole singularities differ is in their sheet location and distance into the complex plane. The single relevant pole for the $a_0$ is located close to the real axis on sheet {\sf IV} (where $\mathrm{Im}\, k_{\pi\eta} > 0, \, \mathrm{Im}\, k_{K\bar{K}} < 0$) while the $f_0$ pole lies further into the complex plane on sheet {\sf II} (${\mathrm{Im}\, k_{\pi\pi} < 0, \, \mathrm{Im}\, k_{K\bar{K}} > 0}$). Considering the similarities in mass and couplings for the $f_0$ and $a_0$ this difference might be considered surprising, but it likely has a relatively simple explanation arising from the only major difference between these two cases: the amount of phase-space for the resonance to decay to the lowest threshold channel, where we have $\rho_{\pi\pi} > \rho_{\pi\eta}$. We can illustrate the effect this has using the simple example of a two-channel Flatt\'e amplitude, in which all elements of the $t$-matrix have a denominator, 
\begin{equation*}
D(s) = m_0^2 -s - i g_1^2\, \rho_1(s) - i g_2^2\, \rho_2(s),
\end{equation*}
where $g_1, g_2$ are real valued couplings to channels labelled $1,2$. In the case of a resonance having a relatively small width, the Flatt\'e amplitude has pole singularities at
\begin{widetext}
\begin{align*}
	\sqrt{s_0} &\approx m_0 \pm \frac{i}{2} \frac{g_2^2\,  \rho_2}{m_0} \left[ \left( \frac{g_1}{g_2} \right)^2 \frac{\rho_1}{\rho_2} - 1 \right] \quad \text{on sheet } \mathsf{II} \text{, if } \; \left( \frac{g_1}{g_2} \right)^2 \frac{\rho_1}{\rho_2} > 1,\,  \text{or}, \\
	\sqrt{s_0} &\approx m_0 \pm \frac{i}{2} \frac{g_2^2\,  \rho_2}{m_0} \left[ 1 - \left( \frac{g_1}{g_2} \right)^2 \frac{\rho_1}{\rho_2} \right] \quad \text{on sheet } \mathsf{IV} \text{, if } \; \left( \frac{g_1}{g_2} \right)^2 \frac{\rho_1}{\rho_2} < 1,\, \text{and}, \\
	\sqrt{s_0} &\approx m_0 \pm \frac{i}{2} \frac{g_2^2\,  \rho_2}{m_0} \left[ 1 + \left( \frac{g_1}{g_2} \right)^2 \frac{\rho_1}{\rho_2} \right] \quad \text{on sheet } \mathsf{III} \text{, in all cases},
\end{align*}
\end{widetext}
where $\rho_i$ is the phase-space for channel $i$ evaluated at $s=m_0^2$. If we use the magnitudes of the couplings we extracted from the residues of poles, as in Eq.~\ref{eq:f0a0_couplings}, as the values $g_1, g_2$, we have for both $a_0$ and $f_0$, $g_1/g_2 \sim 0.8$. However, due to the difference in phase-space, we have $\left( \frac{g_1}{g_2} \right)^2 \!\frac{\rho_1}{\rho_2}$ being slightly less than unity for the $a_0$, but somewhat larger than unity for the $f_0$, explaining the difference in sheet location. In addition this analysis also offers an explanation for the smaller total width of the $a_0$ as being due to the near exact cancellation between terms in $\left[ 1 - \left( \frac{g_1}{g_2} \right)^2 \frac{\rho_1}{\rho_2} \right]$. Within this simple model, there is also an additional sheet {\sf III} pole, but we observe that it will be much further into the complex plane due to the two terms summing in $\left[ 1 + \left( \frac{g_1}{g_2} \right)^2 \frac{\rho_1}{\rho_2} \right]$, which matches our results which suggest that any sheet {\sf III} pole lies far into the complex plane.

As argued in the previous section, the dominance of a single pole close to the $K\overline{K}$ threshold may suggest an association with a $K\overline{K}$ molecular configuration. Given the apparent parallels between the $a_0$ and $f_0$ resonances above, it would appear that at this heavier-than-physical light-quark mass, these states may have a common source, being isovector and isoscalar manifestations of the same $K\overline{K}$ bound system.

From the significant volume dependence of the lowest lying energy level seen in Fig.~\ref{fig:model_spec}, one might conclude that the $\sigma$ could be a physically large object, of size comparable to the lattice volumes used~\cite{Luscher:1985dn}, and therefore that it can be interpreted as a molecular $\pi\pi$ state, rather than a compact object bound by confining forces. Since the $\sigma$ is bound for these values of the quark masses, one can directly apply Weinberg's compositeness criterion~\cite{Weinberg:1962hj} to this state. This approach relates the $S$-wave effective range parameters (${k \cot \delta_0 = a^{-1} + \tfrac{1}{2} r k^2 + \ldots}$) to the probability, $Z$, of finding the $\sigma$ in an elementary bare-particle state,
 \begin{align}
a &= - 2 \frac{1-Z}{2-Z} \frac{1}{\sqrt{m_\pi \epsilon}},  \label{eq:WeinbergA}\\
r &= - \frac{Z}{1-Z} \frac{1}{\sqrt{m_\pi \epsilon}}.  \label{eq:WeinbergR}
\end{align}
where $\epsilon = 2m_\pi - m_\sigma$ is the binding energy of the $\sigma$, and where each equation potentially receives corrections of order of the range of the $\pi\pi$ interaction. If $Z=1$ the $\sigma$ is purely a compact state, while if $Z=0$ it is purely a $\pi\pi$ molecule. Using the values obtained for $a$ and $r$ presented in Section~\ref{Sec:elastic}, and a binding energy determined from the $\sigma$ pole mass, Eqs.~\ref{eq:WeinbergA} and \ref{eq:WeinbergR} give compatible estimates of $Z \sim 0.3(1)$. This suggests that for quark masses where $m_\pi \sim 391\,\mathrm{MeV}$ the $\sigma$ can be understood as being predominantly a $\pi\pi$ molecule. 

While we have successfully determined a complete nonet of scalar mesons with $m_\pi \sim 391$~MeV, it remains to be seen how these states evolve with quark mass. To date, the only one of these states that have been studied using smaller values of the quark mass is the $\sigma$, where in Ref.~\cite{Briceno:2016mjc} it was found that when the pion mass is approximately $236$~MeV, the $\sigma$ has become a broad resonance, resembling somewhat the experimental situation. This is consistent with the expectation that the $\sigma$ should become increasingly unstable as its phase-space for decay to $\pi\pi$ opens up. With this same line of logic, one might also expect the $\kappa$ pole, which at $m_\pi \sim 391\,\mathrm{MeV}$ resides below threshold on the unphysical sheet, to become a complex-valued pole on the second sheet above threshold as the pion mass is decreased. Similarly, one can expect the phase space of the $a_0$ and $f_0$ resonances for decay to the lower channel to increase, and following the discussion above of these poles in the context of the Flatt\'e parametrization, if the decay channel couplings have relatively mild quark-mass dependence, one would expect the poles for $f_0$ and $a_0$ to both come to reside on sheet {\sf II}. 

It is also interesting to consider what would happen to the scalar nonet if the light quark mass were \emph{increased}. For instance, does the $\kappa$ become a real bound state below $\pi K$ threshold on the physical sheet for increasingly heavy quarks as is expected in unitarized chiral perturbation theory (U$\chi$PT)~\cite{Nebreda:2010wv}? What happens if the light quark mass is increased until it is equal to the strange quark mass, meaning the theory has an exact SU(3) flavor symmetry? In this limit there is no splitting between the $\pi\pi, K\overline{K}, \eta \eta$ thresholds, and the scalar nonet is split into an SU(3) octet and a singlet. The evolution of the nonet in this limit has been previously studied using U$\chi$PT~\cite{Oller:2003vf}, where a particular trajectory was chosen which transformed from the physical point to the $m_\pi=m_K=m_\eta=300$~MeV point. A smooth evolution of the $\sigma$ pole to a singlet pole lying well below threshold was observed, while all other (octet) states move to a common complex-value pole on the unphysical sheet.

\vspace{1cm}

The lightest set of tensor mesons determined at this value of the light-quark mass,

\vspace{2mm}\hspace{1cm}
\begin{tabular}{r l l}
$f_2^\mathsf{a}$ 	&\quad $1470(15)$ & $- \tfrac{i}{2} 160(18)\,\mathrm{MeV}$  \\[1.0ex]
$a_2$			 	&\quad $1505(5)$  & $- \tfrac{i}{2} 20(3)\,\mathrm{MeV}$   \\[1.0ex]
$K_2^\star$			&\quad $1577(7)$  & $- \tfrac{i}{2} 66(7)\,\mathrm{MeV}$  \\[1.0ex] 
$f_2^\mathsf{b}$ 	&\quad $1602(10)$ & $- \tfrac{i}{2} 54(14)\,\mathrm{MeV}$,\\[1.0ex]
\end{tabular}

\noindent suggests a rather simple interpretation in terms of \mbox{$q\bar{q}$-like} states, lightly ``dressed'' by their meson-meson decays. The masses and dominant decays
\footnote{see the discussion of the `OZI' rule in the previous section.}
of these states suggest a picture where $f_2^\mathsf{a} \sim \tfrac{1}{\sqrt{2}} \big( u\bar{u} + d\bar{d} \big)$, $a_2  \sim \tfrac{1}{\sqrt{2}} \big( u\bar{u} - d\bar{d} \big)$, $K_2^\star \sim u\bar{s}$, $f_2^\mathsf{b} \sim s\bar{s}$, with negligible flavor mixing in the isoscalar sector. The small mass difference between $f_2^\mathsf{a}$ and $a_2$, despite them being constructed from the same quarks, can be ascribed to the small contribution of disconnected diagrams to the $f_2^\mathsf{a}$ where the quarks annihilate. The larger width of the $f_2^\mathsf{a}$ can be explained by it having allowed decays to $\pi\pi$ with a large phase-space, a channel which is not allowed for the $G$-parity negative isovector $a_2$.

%%%%% SUMMARY %%%%%
\section{Summary \label{Sec:summary}}
% !TEX root = ../f0_paper.tex

%%%%% SUMMARY %%%%    

In this paper, we present the first study with first-principles QCD of low-energy isoscalar $J=0$ and $J=2$ coupled $\pi\pi,K\overline{K},\eta\eta$ scattering amplitudes, and their resonance content. This, in conjunction with previous works~\cite{Briceno:2016mjc, Dudek:2016cru, Dudek:2014qha,Wilson:2014cna}, allows us to paint a complete picture of two different low-lying SU(3) nonets at $m_\pi \sim 400$ MeV. 

The tensor mesons manifest as clear narrow bumps, and from their decay couplings and masses, one can conclude that they behave like a quark-model $q\bar{q}$ nonet.

The scalars prove to be far richer in structure. The $f_0$ and $a_0$ both appear right at $K\overline{K}$ threshold, but they manifest themselves differently -- the $a_0$ appears as an asymmetric peak, while the $f_0$ appears as a dip in a broad enhancement -- the source of this difference can be traced to the interference between the $f_0$ and the $\sigma$, and the absence of a $\sigma$-like state in the isovector channel. Nevertheless, the $f_0$ and $a_0$ are found to have very similar pole properties. The $\sigma$ appears as a bound state, which is likely to be dominated by a $\pi\pi$ molecular configuration, while the $\kappa$ emerges as a virtual bound-state. 

The calculations presented in this paper and in~\cite{Briceno:2016mjc, Dudek:2016cru, Dudek:2014qha,Wilson:2014cna} demonstrate that the finite-volume spectrum approach within lattice QCD can expose non-trivial resonance physics that can significantly inform our understanding of hadron spectroscopy within QCD.

%%%%%%%%%%%%%%%%%%%%%%%%%%%%%%%%%%%%%%%%%%%%%%%%%%%%%%%%%%%%%%%%%%%%%%%%%%%%%%%%%
\begin{acknowledgments}

We thank our colleagues within the Hadron Spectrum Collaboration and additionally M.R. Pennington for fruitful discussions. The software codes {\tt Chroma}~\cite{Edwards:2004sx} and {\tt QUDA}~\cite{Clark:2009wm,Babich:2010mu} were used for the computation of the quark propagators. The contractions were performed on clusters at Jefferson Lab under the USQCD Initiative and the LQCD ARRA project. This research was supported in part under an ALCC award, and used resources of the Oak Ridge Leadership Computing Facility at the Oak Ridge National Laboratory, which is supported by the Office of Science of the U.S. Department of Energy under Contract No. DE-AC05-00OR22725. This research used resources of the National Energy Research Scientific Computing Center (NERSC), a DOE Office of Science User Facility supported by the Office of Science of the U.S. Department of Energy under Contract No. DE-AC02-05CH11231. Gauge configurations were generated using resources awarded from the U.S. Department of Energy INCITE program at Oak Ridge National Lab, and also resources awarded at NERSC. RAB, RGE and JJD acknowledge support from U.S. Department of Energy contract DE-AC05-06OR23177, under which Jefferson Science Associates, LLC, manages and operates Jefferson Lab. JJD acknowledges support from the U.S. Department of Energy Early Career award contract DE-SC0006765. DJW acknowledges funding from the European Union's Horizon 2020 research and innovation programme under grant agreement No 749850 - XXQCD, and was supported by a grant from the Simons Foundation to the Hamilton Mathematics Institute at Trinity College Dublin.

\end{acknowledgments}

%%%%% BIBLIO %%%%%
\bibliographystyle{apsrev4-1}
\bibliography{f0_bib.bib}

\appendix
\widetext
\section{Role of a sheet {\sf III} pole in $S$-wave coupled $\pi\pi,K\overline{K}$ scattering \label{Sec:Jost_scan}}
% !TEX root = ../f0_paper.tex

\begin{figure}[b]
\includegraphics[width = 0.85\textwidth]{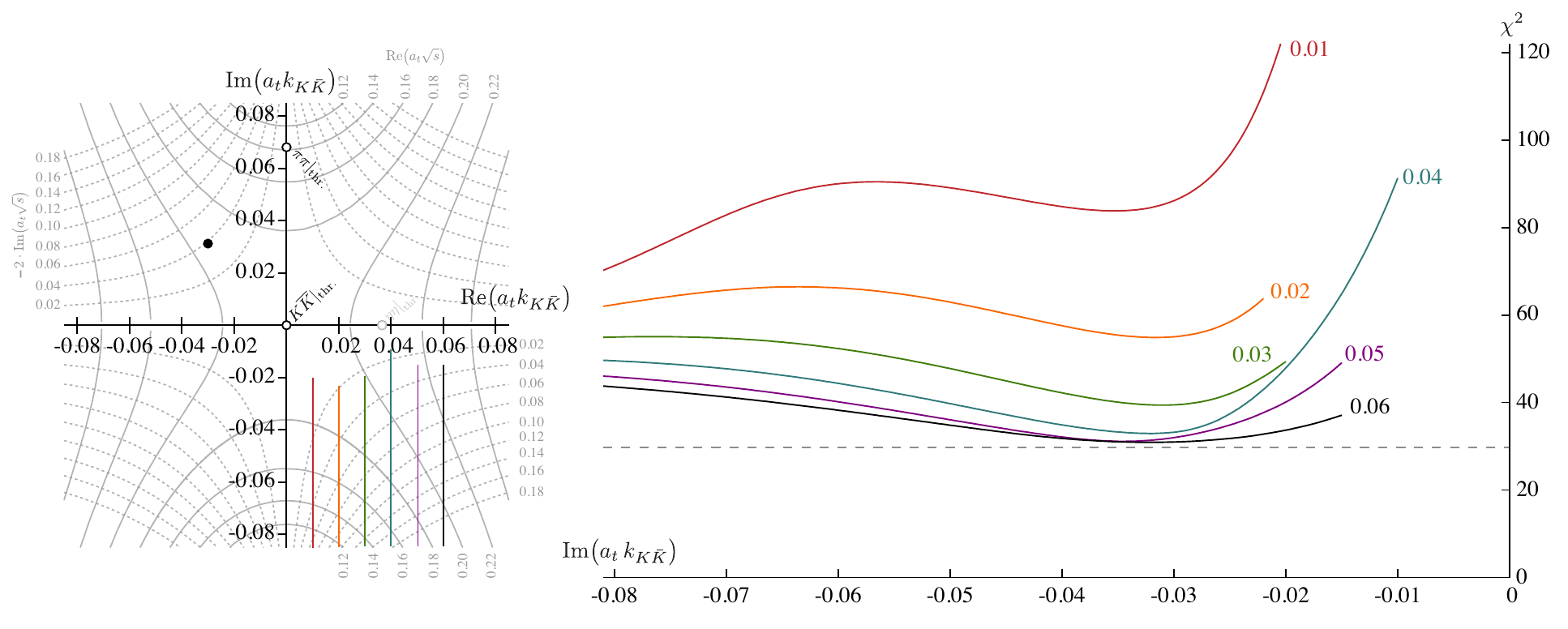}
\includegraphics[width = 0.75\textwidth]{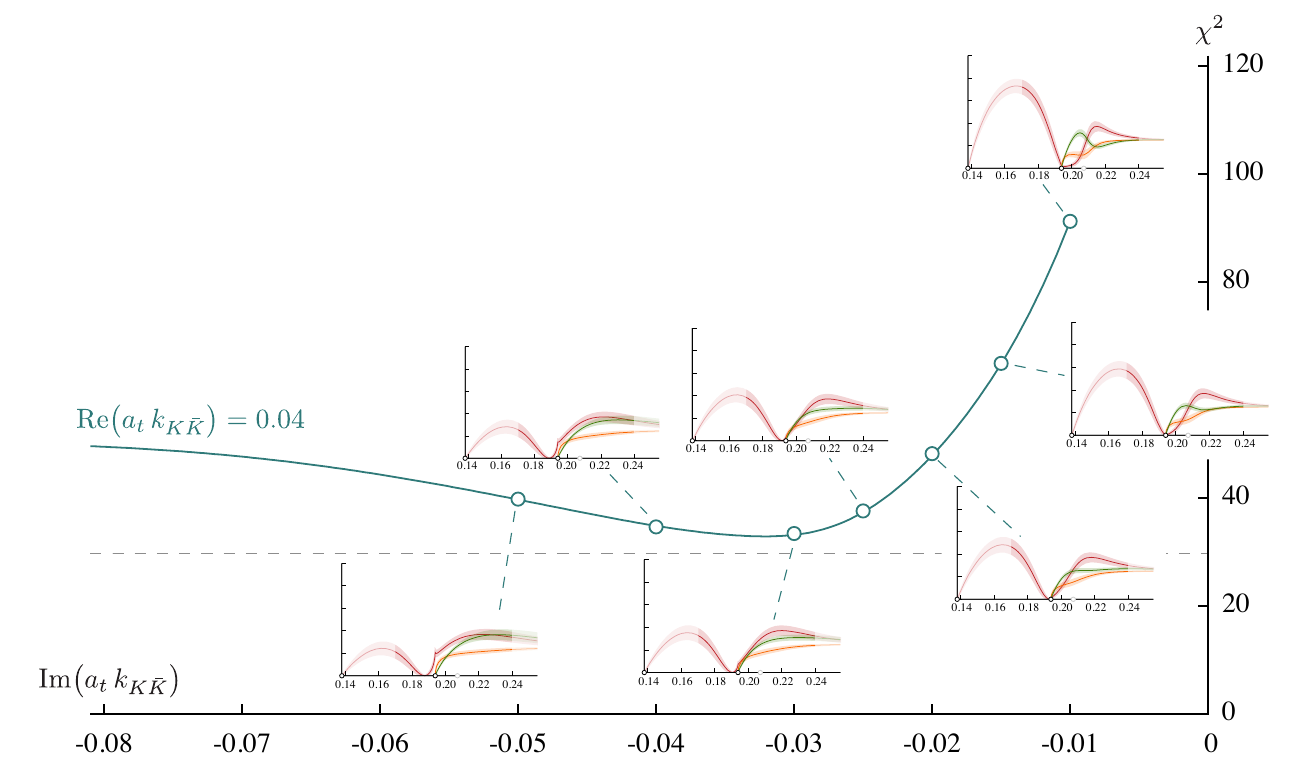}
\caption{With a fixed sheet {\sf II} pole (black circle) (a) scan over possible positions of a sheet {\sf III} pole, allowing the background to adjust each time.  (b) Focus on $\mathrm{Re}( a_t\, k_{K\bar{K}} ) = 0.04$, showing the resulting best-fit amplitudes at several values of $\mathrm{Im}( a_t\, k_{K\bar{K}} )$.  }\label{fig:jost_sheetIII}
\end{figure}

Using the Jost formalism of Section~\ref{sec:Jost}, we can explore the role played by a sheet {\sf III} pole supplementing the required sheet {\sf II} pole describing coupled $\pi\pi, K\overline{K}$ $S$-wave scattering. Fixing a sheet {\sf II} pole at ${a_t \sqrt{s_0} = 0.1941 - \tfrac{i}{2} 0.0389 }$, we have scanned over possible positions for a sheet {\sf III} pole, allowing two parameters in the exponentiated background polynomial to vary at each new position. Figure~\ref{fig:jost_sheetIII}(a) illustrates the resulting variation in the $\chi^2$ describing the lattice QCD spectrum, which indicates that a pole close to physical scattering is disfavored.

%%%%%%%%%%%%%%%%%%%%%%%%%%%%%%%%%%%%%%%%%%%%%%%%%%%%%%%%%%%%%%%%%%%%%
%%%%%%%%%%%%%%%%%%%%%%%%%%%%%%%%%%%%%%%%%%%%%%%%%%%%%%%%%%%%%%%%%%%%%

Figure~\ref{fig:jost_sheetIII}(b), which focusses on the case where $\mathrm{Re}( a_t\, k_{K\bar{K}} ) = 0.04$,  makes clear why a nearby sheet {\sf III} pole is problematic -- it will tend to produce a rapid variation of the amplitudes on the real energy axis which the lattice QCD spectrum is not in agreement with. When a sheet {\sf III} pole is more distant however, its effect can be largely compensated by the background function, and a reasonable $\chi^2$ attained.

We note that attempts to supplement the sheet {\sf II} pole with a second pole located on sheet {\sf IV} lead to very poor descriptions of the spectra or amplitudes which violate the $\big| S_{\pi\pi, \pi\pi} \big| \le 1$ unitarity bound. Our lattice QCD spectrum is clearly incompatible with such a pole distribution.

\end{document}